\documentclass{jfm}

\usepackage{newtxtext}
\usepackage{newtxmath}
\usepackage{mathtools}
\usepackage{xcolor}

\usepackage{natbib}
\usepackage{hyperref}
\usepackage{siunitx}
\usepackage{color,soul}
\hypersetup{
    colorlinks = true,
    urlcolor   = blue,
    citecolor  = blue,
}

\newcommand{\RomanNumeralCaps}[1]
\linenumbers

\newcommand{\x}{\boldsymbol{x}}

\newcommand{\vel}{\boldsymbol{v}}

\newcommand{\nrm}{\boldsymbol{n}}

\newcommand{\grid}[1]{\mathsf{#1}}

\newcommand{\interp}{\prescript{\surface}{}{\grid{E}_\nodes}}
\newcommand{\interpf}{\prescript{\vsurface}{}{\grid{E}_\faces}}
\newcommand{\interpc}{\prescript{\surface}{}{\grid{E}_\centers}}

\newcommand{\gradgrid}{\grid{G}}
\newcommand{\divgrid}{\grid{D}}

\newcommand{\curlgrid}{\grid{C}}
\newcommand{\curlgridsurf}{\grid{C}_\surface}
\newcommand{\lapgrid}{\grid{L}}

\newcommand{\invlapgrid}{\grid{L}^{-1}}

\newcommand{\numpts}{N_s}
\newcommand{\numvorts}{N_v}
\newcommand{\gspace}[1]{\mathcal{#1}}
\newcommand{\faces}{\gspace{F}}
\newcommand{\centers}{\gspace{C}}

\newcommand{\nodes}{\gspace{N}}
\newcommand{\vortices}{\gspace{W}}
\newcommand{\surface}{\gspace{S}}
\newcommand{\vsurface}{\gspace{V}}

\newcommand{\domain}{\Omega}

\newcommand{\onesgrid}{\grid{1}}

\newcommand{\ip}[2]{{\langle #1,#2 \rangle}}
\newcommand{\ipbig}[2]{{\biggl< #1,#2 \biggr>}}

\newcommand{\ipnodes}[2]{\ip{#1}{#2}_{\nodes}}

\newcommand{\spoints}{\surface^{\numpts}}
\newcommand{\vecpoints}{\vsurface^{\numpts}}
\newcommand{\vpoints}{\vortices^{\numvorts}}
\newcommand{\pointwise}[1]{\mathsfi{#1}}
\newcommand{\spoint}[1]{\pointwise{#1}}
\newcommand{\onesspoint}{\spoint{1}}
\newcommand{\unitspoint}[1]{\spoint{e}_{#1}}

\newcommand{\interpv}{\prescript{\vortices}{}{\grid{E}}_\nodes}
\newcommand{\regv}{\prescript{\nodes}{}{\grid{R}}_\vortices}

\newcommand{\regds}{\prescript{\nodes}{}{\grid{R}_\surface}}
\newcommand{\regdssmooth}{\prescript{\nodes}{}{\tilde{\grid{R}}_\surface}}
\newcommand{\regdsf}{\prescript{\faces}{}{\grid{R}_\vsurface}}

\newcommand{\dx}{\Delta x}
\newcommand{\arcparam}{u}
\newcommand{\surfarea}{S}
\newcommand{\dS}{\Delta \surfarea}

\newcommand{\normbase}{n}
\newcommand{\normvecc}[1]{\spoint{\normbase}_{#1}}
\newcommand{\had}{\circ}
\newcommand{\ipscalar}[2]{\ip{#1}{#2}_{\spoints}}
\newcommand{\ipscalarbig}[2]{\ipbig{#1}{#2}_{\spoints}}

\newcommand{\body}{b}
\newcommand{\sfbase}{s}
\newcommand{\vortgrid}{\grid{w}}
\newcommand{\velgrid}{\grid{v}}
\newcommand{\sfgrid}{\grid{\sfbase}}
\newcommand{\potgrid}{\phiup}
\newcommand{\pressgrid}{\grid{p}}
\newcommand{\sfgridinf}{\grid{\sfbase}_{\infty}}
\newcommand{\unitgridvort}[1]{\grid{d}_{#1}}
\newcommand{\vortcirc}[1]{\vortexstrengthq{#1}}

\newcommand{\diagmat}[1]{\grid{D}_{#1}}

\newcommand{\velsurf}{\spoint{v}_{\body}}
\newcommand{\velsurfc}[1]{\spoint{v}_{\body,#1}}
\newcommand{\sfsurf}{\spoint{\sfbase}_{\body}}
\newcommand{\sfsurfp}{\spoint{\sfbase}'_{\body}}
\newcommand{\sfzero}{\spoint{\sfbase}_{0}}
\newcommand{\szero}{\sfbase_{0}}
\newcommand{\szeroj}[1]{\sfbase_{0,#1}}
\newcommand{\vsheet}{\spoint{f}}

\newcommand{\vsheetzero}{\spoint{f}_{0}}
\newcommand{\vsheetzeroj}[1]{\spoint{f}_{0,#1}}
\newcommand{\svsheet}{\tilde{\spoint{f}}}
\newcommand{\unitsvsheet}[1]{\svsheet_{#1}}
\newcommand{\dpsurf}{\Delta\spoint{p}}
\newcommand{\psurf}{\spoint{p}}

\newcommand{\circsurf}{\Gamma_{\body}}
\newcommand{\circzero}{\Gamma_{0}}
\newcommand{\circw}{\Gamma_{\vortgrid}}
\newcommand{\circwj}[1]{\Gamma_{\vortgrid,#1}}

\newcommand{\basicmat}{\mathcal{A}}
\newcommand{\basicBtwo}{\mathcal{B}_{2}}
\newcommand{\basicBoneT}{\mathcal{B}_{1}^{T}}
\newcommand{\basicC}{\mathcal{C}}
\newcommand{\basicS}{\mathcal{S}}
\newcommand{\basicrone}{\mathsfb{r_{1}}}
\newcommand{\basicrtwo}{\mathsfb{r_{2}}}
\newcommand{\basicx}{\mathsfb{x}}
\newcommand{\basicy}{\mathsfb{y}}

\newcommand{\schurA}{\grid{S}}
\newcommand{\schurAsmooth}{\tilde{\grid{S}}}

\newcommand{\projectK}[1]{\grid{P}^{K}_{#1}}
\newcommand{\id}{\grid{I}}
\newcommand{\projectC}{\grid{P}^{\Gamma}}

\newcommand{\sucpar}[1]{\sigma_{#1}}
\newcommand{\sucparmin}[1]{\sigma^{\mathrm{min}}_{#1}}
\newcommand{\sucparmax}[1]{\sigma^{\mathrm{max}}_{#1}}

\newcommand{\fmin}[1]{\tilde{f}^{\mathrm{min}}_{#1}}
\newcommand{\fmax}[1]{\tilde{f}^{\mathrm{max}}_{#1}}

\newcommand{\rsurf}[1]{\spoint{r}_{#1}}
\newcommand{\xsurf}{\rsurf{x}}
\newcommand{\ysurf}{\rsurf{y}}

\newcommand{\xgrid}{\grid{x}}
\newcommand{\ygrid}{\grid{y}}

\newcommand{\linimp}[1]{P_{#1}}
\newcommand{\angimp}[1]{\Pi_{#1}}

\newcommand{\surfb}{\mathcal{S}_\body}
\newcommand{\volb}{\mathcal{V}_\body}
\newcommand{\xcent}{X_c}
\newcommand{\ycent}{Y_c}
\newcommand{\xcentvec}{\boldsymbol{X}_c}

\newcommand{\xvortex}{X}
\newcommand{\yvortex}{Y}

\newcommand{\posvortex}{\boldsymbol{X}}
\newcommand{\vortexstrengths}{\Gamma_{v}}
\newcommand{\vortexstrengthq}[1]{\Gamma_{v,#1}}

\newcommand{\contvortexstrengthq}[1]{\Gamma_{v,#1}}

\newcommand{\vsheetzeroopbase}{\hat{\vsheetzero}}

\newcommand{\vsheetzeroop}{\vsheetzeroopbase^{T}}

\newcommand{\vsheetzeroopj}[1]{\hat{\vsheet}^{T}_{0,#1}}

\usepackage{url}
\usepackage{framed}
\colorlet{shadecolor}{yellow}

\title{Planar potential flow on Cartesian grids}

\author{Diederik Beckers\aff{1}
 \and Jeff D. Eldredge\aff{1}
  \corresp{\email{jdeldre@ucla.edu}}}

\affiliation{\aff{1}Mechanical and Aerospace Engineering, University of California, Los Angeles, CA 90095-1597 USA}

\begin{document}
\maketitle

\begin{abstract}
Potential flow has many applications, including the modelling of unsteady flows in aerodynamics. For these models to work efficiently, it is best to avoid Biot-Savart interactions. This work presents a grid-based treatment of potential flows in two dimensions and its use in a vortex model for simulating unsteady aerodynamic flows. For flows consisting of vortex elements, the treatment follows the vortex-in-cell approach and solves the streamfunction-vorticity Poisson equation on a Cartesian grid after transferring the circulation from the vortices onto the grid. For sources and sinks, an analogous approach can be followed using the scalar potential. The combined velocity field due to vortices, sinks, and sources can then be obtained using the Helmholtz decomposition. In this work, we use several key tools that ensure the approach works on arbitrary geometries, with and without sharp edges. Firstly, the immersed boundary projection method is used to account for bodies in the flow and the resulting body-forcing Lagrange multiplier is identified as the bound vortex sheet strength. Secondly, sharp edges are treated by decomposing the vortex sheet strength into a singular and non-singular part. To enforce the Kutta condition, the non-singular part can then be constrained to remove the singularity introduced by the sharp edge. These constraints and the Poisson equation are formulated as a saddle-point system and solved using the Schur complement method. The lattice Green's function is used to efficiently solve the discrete Poisson equation with unbounded boundary conditions. The method and its accuracy are demonstrated for several problems.
\end{abstract}



\section{Introduction}

Potential flow plays an important role in aerodynamic modelling, but also appears in other areas such as the modelling of water waves or wind farms and the calculation of added mass. Besides its prominent use for steady flow around airfoils at high Reynolds numbers, potential flow theory has long provided the tools for vortex methods to simulate unsteady flows around airfoils and bluff bodies. These vortex methods discretize the vorticity in the flow with singular elements such as point vortices, vortex sheets, or a combination of both. In the case of an inviscid and incompressible model, the irrotational flow outside of these singular vortex elements is a potential flow. Singular vortex elements that represent the free vorticity in an inviscid vortex model are advected by the local flow velocity according to Helmholtz's second theorem and can be tracked as Lagrangian points. In case the vortex method inserts new vortex elements in the flow behind a bluff body or at sharp edges, Kelvin's circulation theorem dictates that the circulation should be conserved. Singular potential flow elements can also serve to enforce the no-penetration condition, with the most common choice in vortex methods being a distribution of singular vorticity on the body, denoted as the bound vortex sheet. Besides their choice for the type of singular elements, potential flow solvers for vortex methods differ in their way of calculating the flow velocity. This can be done by either using direct interaction between the potential flow elements or by calculating the velocity on a grid over the entire domain and the eventual choice dictates the treatment of boundary and edge conditions.

In the first approach, the Green's function of the Laplacian is applied to the Poisson equation in the velocity-vorticity formulation to give the Biot-Savart integral, which provides the exact solution for a velocity field that satisfies unbounded boundary conditions. Biot-Savart vortex methods often smooth the Biot-Savart kernel, equivalent to replacing point vortices by vortex blobs \citep{Chorin1973}, to suppress Kelvin-Helmholtz instabilities below a certain wavelength resulting from the interactions between closely spaced vortex elements. The solution generally requires $\textit{O}(N^2)$ operations, with $N$ the number of vortex elements, to sum the influences of each discretized vortex element on every other element. With fast multipole methods, it scales optimally as $\textit{O}(N)$ but with a large prefactor and overhead cost. Inviscid vortex methods of this kind can straightforwardly use the potential flow tools to enforce the no-penetration, such as conformal mapping or solving the integral equation for a surface singularity distribution through analytical inversion, panel discretization, or with Fourier expansions. For a detailed review of this subject, the reader is referred to \citet{Cottet2000} and \citet{inviscidbook}.

A second approach to calculate the flow velocity follows from the discretization of the Poisson equation in the velocity-vorticity formulation or streamfunction-vorticity formulation on an Eulerian grid over the domain of interest and is called a vortex-in-cell (VIC) approach, first developed by \citet{Christiansen1973}. The procedure requires first to transfer the circulation from Lagrangian  vortex elements onto the grid, then to solve the discrete Poisson equation, and finally to interpolate the velocity (or the curl of the streamfunction) back to the vortex elements. It introduces discretization errors but only requires $\textit{O}(M\log{M})$ operations, with $M$ the number of grid points, to solve the Poisson equation with current numerical techniques and $\textit{O}(N)$ operations to perform the regularization and interpolation. Similar to the regularized Biot-Savart kernel, the grid spacing together with the vorticity regularization scheme determine the cut-off wavelength below which the Kelvin-Helmholtz instabilities get suppressed. 

The early work on VIC methods focused on inviscid vortex dynamics using Fourier-based Poisson solvers with Dirichlet or periodic boundary conditions \citep{Meng1978,Baker1979,Couet1981} and the analysis of different interpolation kernels \citep{Ebania1996}. After viscous schemes for vortex methods were introduced, VIC methods increasingly replaced Biot-Savart methods in an effort to speed up vortex methods for viscous flows, leading to methods with over a billion vortex particles \citep{Chatelain2008}. This also stimulated the development of VIC methods for external flows over bodies, mostly for viscous flows. The most straightforward way to include a body in the flow is to use a body-fitted mesh as in \citet{Cottet2003}, who apply the Helmholtz decomposition on the flow in their VIC method and place a Neumann boundary condition for the scalar potential on the body to account for its presence and employ an analytic boundary condition for the far-field. However, a body-fitted mesh is case-specific and, therefore, the same work (and later also \citet{Poncet2009}), develops an immersed boundary method by introducing a singular distribution of sources that represents the influence of the body and is smeared onto a cartesian grid using a discrete approximation to the Dirac delta function. The result is a source term that is inserted in the Poisson equation for the scalar potential which is solved on the grid. Similar to the immersed boundary methods, Brinkman penalization methods do not require body-fitted meshes. In vortex methods, the Brinkman penalization method \citep{Coquerelle2008,Rossinelli2010,Gazzola2011,Rasmussen2011,Chatelin2014} adds a volume forcing term to the vorticity transport equation that includes a penalization parameter, equivalent to the porosity of the body. However, the method suffers from a strong time step restriction, which motivated \citet{Hejlesen2015} to use an iterative Brinkman penalization method, which \citet{Spietz2017} extended to three dimensions. \citet{Gillis2017} formulates this method as a linear system and uses a recycling iterative solver to obtain the solution more efficiently.

The immersed boundary method and Brinkman penalization method both smear out the influence of the interface onto nearby grid points. \citet{LeVeque1994} developed the immersed interface method (IIM) to overcome this issue and to obtain a higher spatial order of accuracy than the immersed boundary method. The premise of this method is to discretize the jump conditions caused by the interface with finite differences instead of discretizing the Dirac delta function, and the result is a sharp representation of the interface with a second or higher-order accuracy. \citet{Marichal2014} applies the explicit-jump IIM~\citep{Wiegmann2000} in his potential flow method. The influence of the interface was condensed into an extra source term in the streamfunction Poisson equation and it was recognized that the term is equivalent to a bound vortex sheet strength regularized to the grid. The bound vortex sheet strength, streamfunction field, and outer boundary condition on the streamfunction are then computed iteratively. The work presents results from the flow over a cylinder and an airfoil, for which the Kutta condition enforced through discretization of the streamfunction normal derivatives at the trailing edge. \citet{Gillis2018} extends this method, but applies the IIM on the scalar potential instead. An explicit formula for the singular distribution of sources on the interface is then obtained by applying the Sherman-Morrison-Woodbury decomposition formula to the Poisson equation, similar to~\citet{Poncet2009}. Furthermore, by solving the Poisson equation using the lattice Green's function~\citep{katsura71}, which automatically satisfies far-field boundary conditions, the method is no longer iterative and the cost is greatly reduced. \citet{Gillis2019} applies this method again to the streamfunction in two dimensions and employs it in a viscous VIC method.

In this work we present a grid-based treatment for planar potential flow. The focus of this work is on the flow around point vortices, but the treatment can easily be extended to account for sources and sinks. In the case of the flow around vortices, the treatment follows a VIC approach. In our implementation, the streamfunction-vorticity Poisson equation is solved for the streamfunction on the grid using the lattice Green's function, such that unbounded boundary conditions are accounted for. To enforce the no-penetration condition on surfaces in the flow, the treatment is presented by using the immersed-boundary projection method and our implementation is therefore approximately first-order accurate in space~\citep{Colonius2008}. This approach consists in adding an extra singular vorticity source term to the streamfunction-vorticity Poisson equation that is distributed over the discrete surface points and is smeared onto the nearby grid nodes. This extra vorticity term represents the bound vortex sheet strength and assumes the role of a Lagrange multiplier in this method. The modified Poisson equation combined with the no-penetration constraint then forms a saddle-point system that can be solved with the Schur's complement method. Note that the IIM introduces a similar modification to the Poisson equation~\citep{Marichal2014} and the Sherman-Morrison-Woodbury decomposition can produce an expression for the discrete vortex sheet strength~\citep{Gillis2019} that is equivalent to the formula we obtain by using the Schur's compelement method. Consequently, one could use the IIM to obtain a second-order method instead. Drawing inspiration from the analytical treatment of the Kutta condition in Biot-Savart methods, this work then introduces a new way of enforcing the Kutta condition in a discrete potential flow treatment by decomposing the discrete vortex sheet strength into a singular and non-singular part and constraining the non-singular part. This amounts to algebraically constraining the system arising from the immersed-boundary projection method to make it well-behaved.

This paper is structured as follows. We first focus on discretizing the unbounded potential flow problem with point vortices in \S\ref{sec:basicpotentialflow} and then discuss the no-penetration condition in \S\ref{sec:no-penetration}. We describe the role of circulation in \S\ref{sec:non-uniqueness} and introduce the treatment for enforcing the Kutta condition in steady and unsteady flows in \S\ref{sec:kutta}. We extend the treatment to generalized edge conditions in \S\ref{sec:generalized} and introduce methods for computing pressure, impulse, and added mass in \S\ref{sec:force}. The extension to multiple bodies is discussed in \S\ref{sec:multibody}.

\section{Methodology}

\subsection{The basic two-dimensional potential flow problem}
\label{sec:basicpotentialflow}

A two-dimensional, differentiable velocity field $\vel$ on the unbounded domain $\domain = \{ \x=(x,y) \}$ can be decomposed according to the Helmholtz decomposition
\begin{equation} \label{eq:helmholtz}
    \vel= \nabla \phi + \nabla \times \psi \boldsymbol{e}_z,
\end{equation}
where $\phi$ is the scalar potential, $\psi$ is the streamfunction, and $\boldsymbol{e}_z$ is the unit vector out of the plane. In a potential flow without sources or sinks, this problem can be solved by solving exclusively the governing equation for the streamfunction
\begin{equation} \label{eq:psigovequation}
    \nabla^2 \psi = -\omega,
\end{equation}
where $\omega$ is the vorticity field, consisting solely of $\numvorts$ singular point vortices:
\begin{equation} \label{eq:vorticity}
    \omega = \sum_{q=1}^{\numvorts} \contvortexstrengthq{q} \boldsymbol{\delta}(\x - \posvortex_q),
\end{equation}
where $\boldsymbol{\delta}$ is the two-dimensional Dirac delta function, and $\posvortex_q = (\xvortex_q,\yvortex_q)$ and $\contvortexstrengthq{q}$ are the position and strength of the $q$th point vortex.

Because (\ref{eq:psigovequation}) does not account for sources and sinks, one would have to solve the auxiliary Poisson problem for the scalar potential:
\begin{equation} \label{eq:phigovequation}
    \nabla^2 \phi = \Theta,
\end{equation}
where $\Theta$ is the rate of dilatation, consisting of singular sources and sinks, similar to (\ref{eq:vorticity}). The overall velocity field due to vortices, sinks, and sources can then be obtained using the Helmholtz decomposition~(\ref{eq:helmholtz}).

\begin{figure}
    \centering
    \includegraphics[page=1]{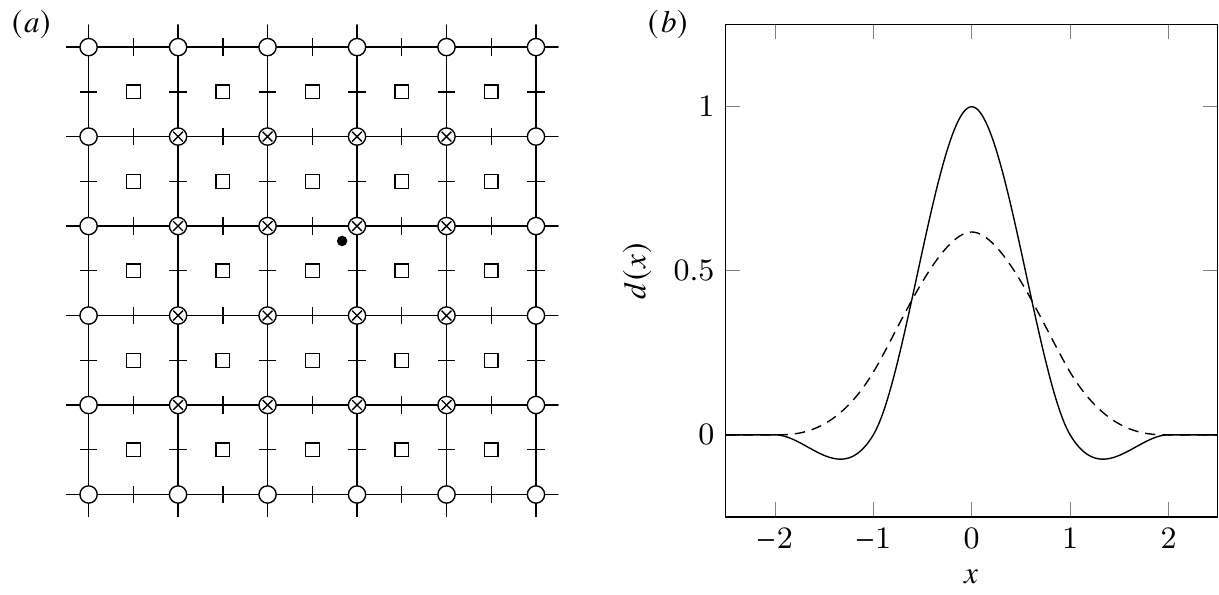}
    \caption{(\textit{a}) Schematic of the grid and the regularization of data from a Lagrangian point (\fullcirc) onto the grid with a discrete Dirac delta function of radius two. Symbols (\opencirc), (\opensquare), (|), and (-) denote the locations holding the components of the nodes $\nodes$, cell centers $\centers$, horizontal faces $\faces_x$, and vertical faces $\faces_y$, respectively. Symbols ($\times$) denote the nodes that are affected by the regularization. (\textit{b}) Two examples of discrete Dirac delta functions of radius two: the $M_4'$ function (\full) from~\citet{Monaghan1985a} and the smoothed three-point function (\dashed) from \citet{Yang2009}.}
    \label{fig:grid_M4}
\end{figure}

We now introduce a discrete treatment of this basic potential flow problem and focus on the discrete streamfunction. We consider here a staggered, Cartesian grid with uniform cell size $\dx$ and of infinite extent. The space corresponding to data at cell vertices (nodes) on this grid is denoted by $\nodes$, and the physical coordinates of these nodes by $\xgrid$ and $\ygrid$. Furthermore, we consider a finite number $\numvorts$ of Lagrangian, singular point vortices. The space of scalar data on these points is denoted by $\vpoints$ and we define $\vortexstrengths \in \vpoints$ as the vector containing the strengths of the point vortices in our grid-based treatment. The basic (unbounded) potential flow problem is expressed as 
\begin{equation}
\label{eq:poisson}
\lapgrid \sfgrid = -\vortgrid,
\end{equation}
where $\lapgrid$ is the discrete 5-point Laplacian operator, $\sfgrid \in \nodes$ is the discrete streamfunction, and $\vortgrid \in \nodes$ the discrete vorticity. The discrete velocity field $\velgrid$, whose components lie on the faces of the cells with the corresponding normals, is computed from $\sfgrid$ by the discrete curl operation,
\begin{equation}
\label{eq:velsf}
    \velgrid = \curlgrid\sfgrid.
\end{equation}
The operator $\curlgrid$ applies centered differences between the nodes to obtain the velocity components at the intermediate centers of cell faces. We denote the space of data that lie on cell faces by $\faces$, so $\curlgrid:\nodes \mapsto \faces$. Figure~\ref{fig:grid_M4}(a) shows the staggered grid structure.

The grid differencing operators $\lapgrid$ and $\curlgrid$ and others to be defined are scaled by the grid spacing, and thus represent second-order approximations of the corresponding continuous operators, and the discrete streamfunction $\sfgrid$, vorticity $\vortgrid$, and velocity $\velgrid$ are each approximations of their continuous counterparts. The total flow circulation is equal to the sum of grid vorticity multiplied by the cell area $\dx^2$. To support the work that follows, we define an inner product on the grid nodes,
\begin{equation}
    \ipnodes{\vortgrid_1}{\vortgrid_2} = \dx^2\vortgrid_1^T \vortgrid_2,
\end{equation}
for $\vortgrid_1$, $\vortgrid_2 \in \nodes$. The total circulation can then be written compactly as
\begin{equation}
    \Gamma_\vortgrid = \ipnodes{\onesgrid}{\vortgrid},
\end{equation}
where $\onesgrid \in \nodes$ is a grid vector of ones.

The particular solution of equation (\ref{eq:poisson}) can be written down immediately with the help of the lattice Green's function for $\lapgrid$ \citep{katsura71,cserti00,liska14}. We denote this simply by the inverse operator,
\begin{equation}
\sfgrid = -\invlapgrid \vortgrid.
\end{equation}
It can be shown that, with a suitable truncation of the grid, both $\lapgrid$ and its inverse are symmetric operators. However, $\lapgrid$ is only positive semi-definite, and an additional homogeneous solution, $\sfgridinf \in \nodes$---for example, corresponding to a uniform flow---can be added to the particular solution of this equation $-\invlapgrid \vortgrid + \sfgridinf$. This homogeneous solution allows us to satisfy boundary conditions at (discrete) infinity. The size of the domain in our simulations is therefore not relevant, as long as it includes the features that are of interest.

As in the vortex-in-cell approach \citep{Christiansen1973}, the discrete vorticity is obtained by immersing the vortex elements into the grid and transferring their circulation to the nearby nodes using a tensor product of two one-dimensional, discrete Dirac delta functions:
\begin{equation}
\label{eq:vortexregularization}
    \vortgrid = \frac{1}{\dx^2}\sum_{q=1}^{\numvorts} \vortexstrengthq{q} d\left(\frac{\xgrid-\xvortex_{q}}{\dx}\right)d\left(\frac{\ygrid-\yvortex_{q}}{\dx}\right),
\end{equation}
where $d$ is a discrete Dirac delta function\footnote{In particular, $d(x/\dx)/\dx$ represents a Dirac sequence as $\dx \rightarrow 0$.}. By the properties imposed on $d$,
\begin{equation}
    \ipnodes{\onesgrid}{\vortgrid}  = \sum_{q=1}^{\numvorts} \vortexstrengthq{q}.
\end{equation}

In this work, $d$ is the $M_4'$ function from~\citet{Monaghan1985a}, depicted in Figure~\ref{fig:grid_M4}(b); panel (a) depicts how data from a Lagrangian point is regularized onto the grid nodes. For compact notation, we define $\unitgridvort{q} \in \nodes$, the grid vorticity field generated when a vortex of unit strength is immersed into the grid, so that we can write (\ref{eq:vortexregularization}) as
\begin{equation}
\label{eq:vortexregularizationsortacompact}
\vortgrid = \sum_{q=1}^{\numvorts} \vortexstrengthq{q} \unitgridvort{q}.
\end{equation}
Even more compactly, we define the {\em regularization operator} $\regv: \vpoints \mapsto \nodes$, whose $\numvorts$ columns are $\unitgridvort{q}$, $q = 1,\ldots,\numvorts$, which allows us to write (\ref{eq:vortexregularizationsortacompact}) as
\begin{equation}
\label{eq:vortexregularizationcompact}
    \vortgrid = \regv \vortexstrengths,
\end{equation}
the discrete streamfunction field as $\sfgrid = -\invlapgrid \regv \vortexstrengths + \sfgridinf$ and the velocity field as
\begin{equation}
    \velgrid = -\curlgrid \invlapgrid \regv \vortexstrengths + \curlgrid \sfgridinf.
\end{equation}

\begin{figure}
    \centering
    \includegraphics[page=1]{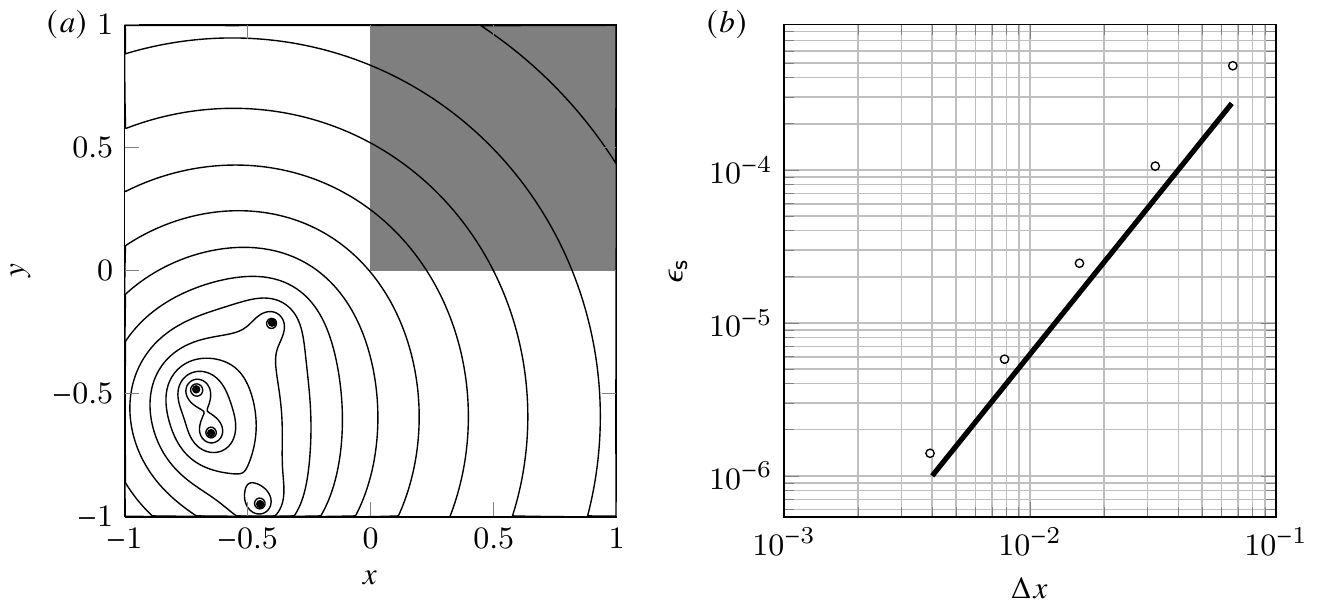}
    \caption{(\textit{a}) Contours (\full) of the discrete streamfunction for randomly positioned point vortices (\fullcirc) of random strengths, and (\textit{b}) its error  (\opencirc) over the shaded area for different grid spacings. Overlaid is an error (\textbf{\full}) that scales as $\Delta x^2$.}
    \label{fig:basicpotentialflow}
\end{figure}

Figure~\ref{fig:basicpotentialflow} shows a spatial grid refinement analysis for a flow consisting of point vortices of random strength that are randomly positioned in the lower-left quadrant of the domain. The analysis verifies that the discretization technique of the Poisson equation is second-order accurate in $\dx$. We compute the error as $\epsilon_{\sfgrid} = \Vert \sfgrid - \psi(\xgrid,\ygrid) \Vert_2 / \Vert \psi(\xgrid,\ygrid) \Vert_2$, where $\psi$ is the exact solution for the streamfunction. We only consider the values in the upper right quadrant of the domain to exclude the positions of the point vortices, because the exact singularities at these positions are not comparable to the regularized, discrete version.

To obtain the velocity at the locations of the point vortices, the discrete velocity field $\velgrid$ should first be interpolated from cell faces to the nodes (using simple averages). Then, the velocity can be interpolated onto the point vortices with the {\em interpolation operator} $\interpv: \nodes\mapsto\vpoints$, which is the transpose of the regularization operator, $\interpv = \regv^{T}$, to obtain an overall interpolation scheme that is consistent with (\ref{eq:vortexregularizationcompact}).

It is worth noting here that, if we wish to include sources and sinks, then we can make additional use of a discrete scalar potential, lying at cell centers, $\potgrid \in \centers$, whose governing equation is the discretized version of (\ref{eq:phigovequation}), analogous to (\ref{eq:poisson}). The sources and sinks can be immersed into a rate of dilatation field at cell centers in similar manner to vortices to cell nodes (\ref{eq:vortexregularization}). The velocity field due to the combination of vortices, sinks, and sources can then be composed using the discrete version of the Helmoltz decomposition ($\ref{eq:helmholtz}$):
\begin{equation}
    \velgrid = \gradgrid\potgrid + \curlgrid\sfgrid.
\end{equation} 
Analogous to the discrete curl operator, the discrete gradient operator $\gradgrid$ applies centered differences to scalar data at cell centers $\centers$ to obtain the velocity components at the cell faces, so $\gradgrid:\centers \mapsto \faces$.

\subsection{Potential flow with an impenetrable surface}
\label{sec:no-penetration}
Now, let us suppose we have a rigid impenetrable surface $\surface$, on which we seek to enforce the no-penetration condition for the streamfunction. The no-penetration condition asserts that the normal components of the fluid velocity and this surface velocity must be equal. For rigid bodies, this surface motion can be alternatively described by a streamfunction, and the no-penetration condition in the absence of sources and sinks can be imposed equivalently (in two dimensions) by setting the fluid streamfunction equal to that of the surface $\psi_b$ up to a uniform value. In continuous form, this is described by the Dirichlet problem
\begin{align} 
    \nabla^2 \psi &= - \omega \label{eq:dirichletproblem1}\\
    \psi \left( \x \right) &= \psi_b \left( \x \right), \, \x \in \surface. \label{eq:dirichletproblem2}
\end{align}
In the presence of sources and sinks, one can again solve the auxiliary problem for the scalar potential and enforce the no-penetration condition only for the flow due to the sources and sinks. The overall velocity field~(\ref{eq:helmholtz}) then satisfies the no-penetration condition. Note that the boundary value problem for the scalar potential is a Neumann problem and requires a slightly different numerical treatment~\citep{Poncet2009,Gillis2018}.
 
The continuous Dirichlet problem~(\ref{eq:dirichletproblem1})--(\ref{eq:dirichletproblem2}) can be solved via Green's theorem with boundary integrals. We will now solve the discrete version of this equation using the immersed boundary projection method, resulting in completely analogous operations.

\subsubsection{Discrete surface and its immersion in the grid}
We enforce the no-penetration condition at a finite number $\numpts$ of discrete surface forcing points; the space of scalar data on these Lagrangian points is denoted by $\spoints$. In particular, let us define $\xsurf, \ysurf \in \spoints$ as the vectors of $x$ and $y$ coordinates of the surface points. Each surface point $p$ is associated with a small straight segment of length $\dS_p$. Some of the calculations will require information about the local surface orientation. For this purpose, we define vectors $\normvecc{x}, \normvecc{y} \in \spoints$ of components of the discrete surface unit normals. We also define the space $\vecpoints$ to hold vector-valued data, such as velocity, on the immersed surface points. For convenience, let us also define unit vectors $\unitspoint{p}$ on this space, equal to 1 at surface point $p$ ($1 \leq p \leq \numpts$) and zero at every other point. For example, the $x$ coordinate of point $p$ is picked out of the vector $\xsurf$ by projection onto the $p$th unit vector:
\begin{equation}
    \unitspoint{p}^T\xsurf.
\end{equation}
Each of the surface point spaces has an associated inner product that includes the surface length, e.g.,
\begin{equation}
    \ipscalar{\spoint{s}_1}{\spoint{s}_2} = \sum_p \dS_p s_{1,p} s_{2,p},  
\end{equation}
for any $\spoint{s}_1, \spoint{s}_2 \in \spoints$, so that the inner product approximates a surface integral. Another vector we will make substantial use of in this paper is $\onesspoint \in \spoints$, a vector of ones on all surface points.

From any vector $\spoint{s} \in \spoints$, we can also form a diagonal $\numpts\times \numpts$ operator $\diagmat{\spoint{s}}$ with the entries of the vector along the diagonal. When this operator acts upon another vector $\spoint{u}\in\spoints$, it represents the Hadamard (i.e., element-by-element) product of the two vectors, $\diagmat{\spoint{s}} \spoint{u} = \spoint{s}\had\spoint{u} \in \spoints$. Note that $\diagmat{\spoint{s}}\spoint{u} = \diagmat{\spoint{u}} \spoint{s}$, and that $\diagmat{\spoint{s}}\onesspoint = \spoint{s}$.

Similar to (\ref{eq:vortexregularizationcompact}), surface data are immersed into the grid with the {\em regularization operator} $\regds: \spoints \mapsto \nodes$. Grid data are interpolated onto the surface points with the {\em interpolation operator} $\interp: \nodes\mapsto\spoints$. $\regds$ can be constructed (and we will assume it has) so that it is the transpose of the interpolation operator, $\regds = \interp^{T}$, with respect to the grid and surface inner products defined in this paper. Furthermore, note that $\regds$ and $\interp$ can be constructed with a different choice for the discrete Dirac delta function than the one used for the vortex regularization. In this work, we use the smoothed three-point function from \citet{Yang2009} (figure~\ref{fig:grid_M4}) and we use a uniform spacing between the surface points.

\subsubsection{The immersed surface potential flow problem}

The surface's motion is specified by a velocity distribution $\vel_{\body}$, represented discretely by components $\velsurfc{x}, \velsurfc{y} \in \spoints$. For rigid bodies, this surface motion can be described by a streamfunction. Specifically, translation at velocity $(U,V)$ and rotation at angular velocity $\Omega$ would be described equivalently by velocity components
\begin{equation}
\label{eq:velsurf}
    \velsurfc{x} = U \onesspoint - \Omega \ysurf, \qquad \velsurfc{y} = V \onesspoint + \Omega \xsurf
\end{equation}
or by a surface streamfunction $\sfsurf \in \spoints$:
\begin{equation}
\label{eq:sfsurf}
    \sfsurf = U\ysurf - V\xsurf -\frac{1}{2}\Omega \left(\diagmat{\xsurf}\xsurf + \diagmat{\ysurf}\ysurf\right).
\end{equation}

The no-penetration condition can be imposed by setting the discrete streamfunction equal to that of the surface, up to a uniform value, $\sfzero \in \spoints$: 
\begin{equation}
\label{eq:nopen}
\interp \sfgrid = \sfsurf - \interp \sfgridinf - \sfzero.
\end{equation}
For later shorthand, we will denote the difference between the body motion streamfunction and interpolated uniform flow streamfunction by $\sfsurfp \equiv \sfsurf - \interp \sfgridinf$. This modified streamfunction simply consists of subtracting the components $(U_\infty,V_\infty)$ of the uniform flow from $(U,V)$ in (\ref{eq:sfsurf}). The uniform value $\sfzero$ is left unspecified and will later serve the role of enforcing a constraint on circulation. For now, we will suppose that it can be set arbitrarily. 

The no-penetration constraint is enforced in the basic potential flow problem (\ref{eq:poisson}) with the help of a vector of Lagrange multipliers, $\vsheet \in \spoints$, on the surface points. The modified potential flow problem is thus
\begin{equation}
\lapgrid \sfgrid = -\left(\vortgrid + \regds \vsheet \right).
\end{equation}

In fact, by simple comparison with the vorticity $\vortgrid$, it is clear that the vector $\vsheet$ represents the strength of the discrete bound vortex sheet on the surface and serves as another source term of the Poisson equation. Suppose we consider the bound vortex sheet $\gamma(\arcparam)$ that emerges from the analogous continuous problem on the undiscretized surface, where $\arcparam$ is the arc-length parameter along the surface. At each point $p$, the discrete solution $\vsheet$ is approximately equal to this continuous solution:
\begin{equation}
    \unitspoint{p}^{T}\vsheet \simeq \gamma(\arcparam_p).
\end{equation}

The vector of Lagrange multipliers $\vsheet$ is initially unknown. Thus, the potential flow problem in the presence of the impenetrable surface is
\begin{equation}
\label{eq:basicblock}
\begin{bmatrix}
\lapgrid & \regds \\
\interp &  0
\end{bmatrix} \begin{pmatrix} \sfgrid \\ \vsheet \end{pmatrix} =
\begin{pmatrix} -\vortgrid \\ \sfsurfp - \sfzero \end{pmatrix}.
\end{equation}
This problem (\ref{eq:basicblock}) has the structure of a generic {\em saddle-point problem} \citep{benzi2005}. We will encounter many such systems in this work, so in appendix \ref{app:saddlesystems} we summarize a solution approach, based on block-LU decomposition.
The generated solution algorithm of (\ref{eq:basicblock}) is
\begin{align}
\lapgrid \sfgrid^{*} &= -\vortgrid \\
\schurA \vsheet &= \sfsurfp - \sfzero  - \interp \sfgrid^{*} \label{eq:vsheet}\\
\sfgrid &= \sfgrid^{*} - \invlapgrid \regds \vsheet,
\end{align}
where the Schur complement $\schurA$ is
\begin{equation}
\schurA = -\interp \lapgrid^{-1} \regds.
\end{equation}

Based on the properties of the matrices comprising $\schurA$, this operator is symmetric and negative definite, and therefore invertible. Its inverse $\schurA^{-1}$, also symmetric, maps a surface distribution of streamfunction to a corresponding bound vortex sheet strength. Note that the computation of $\vsheet$ through $\schurA^{-1}$ is sensitive to both the ratio of discrete surface spacing to the grid spacing $\dS/\dx$ and the choice of discrete Dirac delta function. On the one hand, small values for $\dS/\dx$ and discrete Dirac delta functions with small support generally both lead to more high-frequency noise in $\vsheet$. This is because the underlying continuous problem for $\schurA^{-1}$ is a Fredholm integral equation of the first kind, which is ill-posed. As a result, the discrete analogue problem is poorly-conditioned~\citep{Goza2016}. On the other hand, values for $\dS/\dx$ that are too high can lead to streamlines penetrating a surface. We found that values between one to four can provide a good balance between the smoothness of $\vsheet$ and the accuracy of the streamlines near a surface.

We can describe this algorithm in words: First, solve for the intermediate streamfunction field, associated with vorticity in the fluid, but without regard for the presence of the surface. Second, find the bound vortex sheet whose associated streamfunction cancels the difference between the specified streamfunction on the surface and the intermediate streamfunction evaluated on the surface. Finally, correct the intermediate streamfunction field for the influence of the bound vortex sheet. A version of the Julia code that implements this algorithm, as well as the algorithms in the following sections, is available in the authors' Github repository \citep{gridpotentialflow}.

\begin{figure}
    \centering
    \includegraphics[page=1]{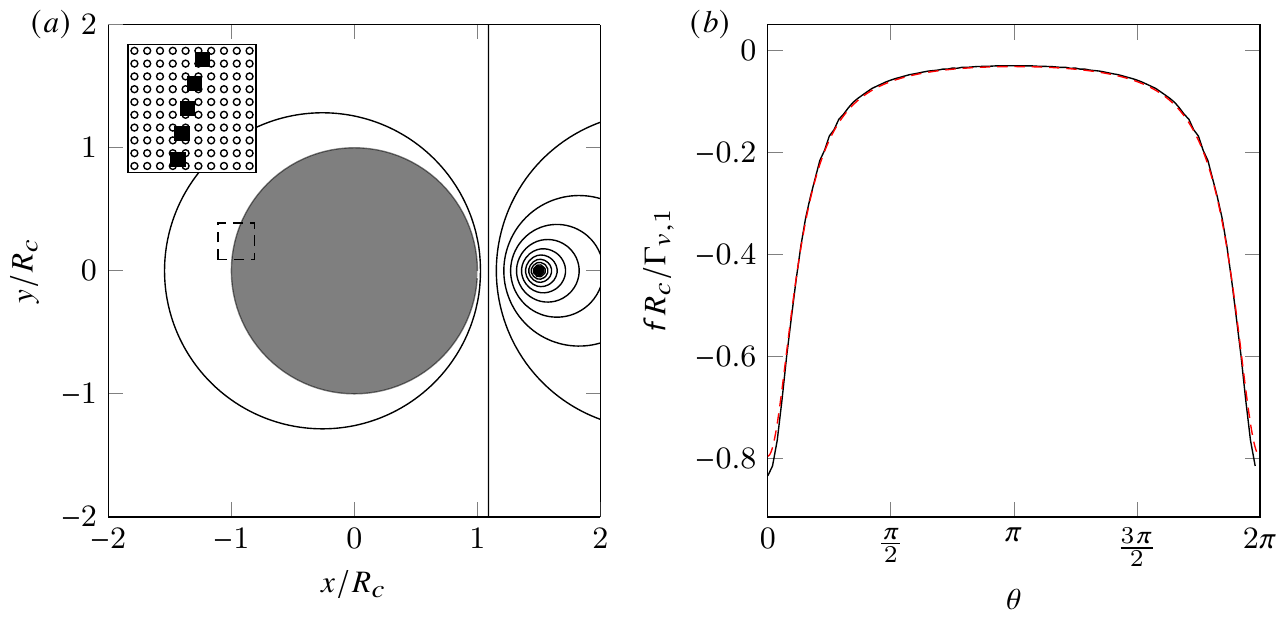}
    \caption{(\textit{a}) Contours (\full) of the discrete streamfunction for a point vortex (\fullcirc) with strength $\vortexstrengthq{1}$ at $(R_v,0)$ near a circular cylinder of radius $R_c$ and with a bound circulation $-\vortexstrengthq{1}$. The inset figure shows a closeup of the nodes (\opencirc) and surface points (\fullsquare) in the boxed area (\dashed). (\textit{b}) The scaled discrete vortex sheet strength (\full) as a function of the angle $\theta$ measured counter-clockwise from the positive $x$-axis. $k=1$ corresponds to the right-most point on the surface and increases counterclockwise. Overlaid is the exact continuous solution  (\textcolor{red}{\dashed}). The simulation is performed with $R_v/R_c = 3/2$, $\dx/R_c=0.03$, and $\dS/\dx=2$.}
    \label{fig:vortexnearcylinder}
\end{figure}

\begin{figure}
    \centering
    \includegraphics[page=1]{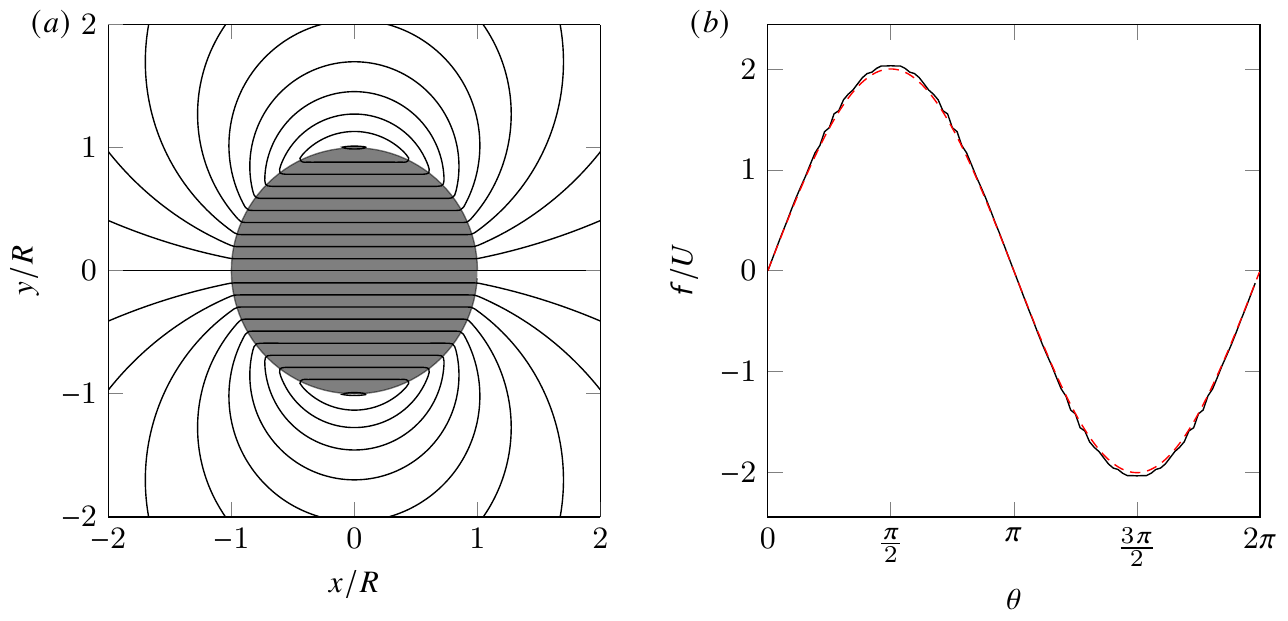}
    \caption{(\textit{a}) Contours (\full) of the discrete streamfunction for a horizontally translating circular cylinder with radius $R$, and (\textit{b}) its scaled discrete vortex sheet strength (\full). Overlaid is the exact continuous solution  (\textcolor{red}{\dashed}). The simulation is performed with $\dx/R=0.03$ and $\dS/\dx=2$.}
    \label{fig:movingcylinder}
\end{figure}

We give two examples of the streamfunction with a body present and show the associated vortex sheet strength. Figure~\ref{fig:vortexnearcylinder} shows a vortex near a circular cylinder and figure~\ref{fig:movingcylinder} shows a circular cylinder that translates horizontally. In both cases, the vortex sheet strength is in good agreement with the analytical solution from potential flow theory.

For later use, we note that the solution of (\ref{eq:basicblock}) can also be written in inverse form using equation (\ref{eq:saddleinverse}):
\begin{equation}
\label{eq:Ainverse}
\begin{pmatrix} \sfgrid \\ \vsheet \end{pmatrix} = \begin{bmatrix} \invlapgrid + \invlapgrid \regds\schurA^{-1} \interp \invlapgrid & -\invlapgrid\regds \schurA^{-1} \\ -\schurA^{-1} \interp\invlapgrid & \schurA^{-1}\end{bmatrix} \begin{pmatrix} -\vortgrid \\ \sfsurfp - \sfzero \end{pmatrix}. 
\end{equation}
The matrix operator in (\ref{eq:Ainverse}) is the inverse of the basic saddle-point system.


\subsection{Non-uniqueness and discrete circulation}
\label{sec:non-uniqueness}
In two-dimensional potential flows, there is no unique solution to problem (\ref{eq:basicblock}), since one can choose any value for the uniform value $\sfzero$ and still enforce the no-penetration condition. Equivalently, we can specify any {\em circulation} about the body and still enforce this condition. Let us determine the relationship between $\sfzero$ and circulation. For later use, let us write this uniform surface streamfunction as $\sfzero = \szero\onesspoint$, where $\szero$ is a single scalar value. The {\em discrete circulation} $\circsurf$ about the body is given by the sum of the bound vortex sheet data and can be written compactly as
\begin{equation}
\label{eq:circdef}
\circsurf = \ipscalar{\onesspoint}{\vsheet}.
\end{equation}
The discrete circulation of the vortex sheet in the solution (\ref{eq:vsheet}) is
\begin{equation}
\label{eq:circs0}
\circsurf = \ipscalarbig{\onesspoint}{\schurA^{-1} \left( \sfsurfp + \interp \invlapgrid \vortgrid\right)}  - \szero \ipscalarbig{\onesspoint}{\schurA^{-1} \onesspoint}.
\end{equation}
Note that we can obtain the same expression if we would use $\szero$ as a Lagrange multiplier to enforce the constraint (\ref{eq:circdef}) as
\begin{equation}
\label{eq:circsystem}
\begin{bmatrix}
\lapgrid & \regds & 0 \\
\interp & 0 & \onesspoint \\
0 & \onesspoint^{T} \diagmat{\dS}  & 0
\end{bmatrix}
\begin{pmatrix}
\sfgrid \\
\vsheet \\
\szero
\end{pmatrix} = 
\begin{pmatrix}
-\vortgrid \\
\sfsurfp  \\
\circsurf
\end{pmatrix},
\end{equation}
where we use the fact that the inner product $\ipscalar{\onesspoint}{\vsheet}$ can, by its definition, be rewritten as $\onesspoint^T \diagmat{\dS} \vsheet$. Here, $\diagmat{\dS}$ is a diagonal matrix containing the surface element arc lengths.We will use $\szero$ in a similar way in the next section to enforce the Kutta condition and demonstrate how to solve the associated saddle-point system.

The scalar factor $\ipscalar{\onesspoint}{\schurA^{-1} \onesspoint}$ in expression (\ref{eq:circs0}) is a property of the set of points and their immersion into the Cartesian grid. Part of this factor, $\schurA^{-1} \onesspoint$, represents the bound vortex sheet strength associated with a uniform, unit-strength streamfunction on the surface. This sheet has a particularly important role in some of the discussion to follow, so we will denote its strength by $\vsheetzero$:
\begin{equation}
\label{eq:f0}
\vsheetzero \equiv \schurA^{-1} \onesspoint.
\end{equation}
The transpose of $\vsheetzero$, equal to $\onesspoint^{T}\schurA^{-1}$, calculates the circulation of the associated bound vortex sheet when it acts upon a surface streamfunction. Thus, the factor $\ipscalar{\onesspoint}{\schurA^{-1} \onesspoint}$ is the circulation associated with a uniform, unit-strength surface streamfunction. We will refer to this as $\circzero$:
\begin{equation}
\label{eq:circ0}
\circzero \equiv \ipscalar{\onesspoint}{\schurA^{-1} \onesspoint} \equiv \ipscalar{\onesspoint }{\vsheetzero} \equiv\ipscalar{\vsheetzero}{\onesspoint}. 
\end{equation}
We can rewrite the inner product $\ipscalar{\vsheetzero}{\onesspoint}$ as  $\vsheetzero^T \diagmat{\dS} \onesspoint$, and we will define $\vsheetzeroopbase = \diagmat{\dS} \vsheetzero$, for shorthand in what follows. By this notation, $\circzero = \vsheetzeroop \onesspoint$, and $\vsheetzeroop$ applied to any surface streamfunction obtains the corresponding circulation.

\begin{figure}
    \centering
    \includegraphics[page=1]{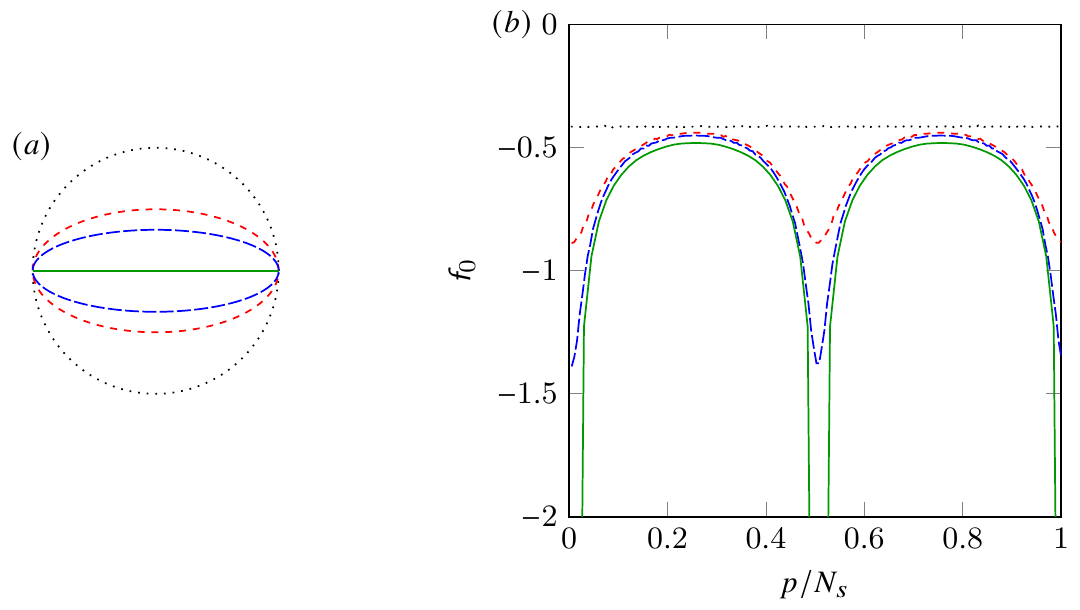}
    \caption{(\textit{a}) Geometry and (\textit{b}) the scaled bound vortex sheet strength $\vsheetzero$, associated with a uniform, unit-strength streamfunction, as a function of the surface point index $p$ divided by the number of surface points $\numpts$ for elliptical cylinders with different aspect ratios ($AR$): \dotted, $AR = 1$; \textcolor{red}{\dashed}, $AR = 2$; \textcolor{blue}{\broken}, $AR = 3$; \textcolor{green!60!black}{\full}, $AR = \infty$ (flat plate). The simulations are performed with $\dx/R=0.004$ and $\dS/\dx=3$.}
    \label{fig:aspectratios}
\end{figure}

Figure~\ref{fig:aspectratios} shows the distribution of $\vsheetzero$ for elliptical cylinders with different aspect ratios. For a circular cylinder, the flow due to a uniform streamfunction on the body corresponds to the flow when the cylinder is replaced by a point vortex at its center. The streamlines are concentric circles and the tangential velocity is constant at a given radius. The resulting vortex sheet strength $\vsheetzero$ therefore assumes a uniform distribution. The figure also demonstrates that when the aspect ratio increases, the distribution gradually shows stronger variations near the edges of the major axis, corresponding to an acceleration and deceleration of the flow when it passes those edges. The distribution eventually turns singular at the edges of a flat plate as the flow now has to navigate around a sharp corner, which will be discussed in more detail in the next section. To clearly illustrate the emergence of the singularities, the simulations in this figure are performed with an extremely fine grid ($\dx/R=0.004$). Note that such fine grids are not needed in general, as is demonstrated by the other figures in this work.

The last term in (\ref{eq:circs0}) illustrates the direct relationship between the scalar value $\szero$ and the bound circulation $\circsurf$, and we identified $\szero$ as a means of setting the circulation. Before we use it in the next section to enforce the Kutta condition, we will use it here for an immediate purpose. The prescribed surface streamfunction $\sfsurfp$ (given by (\ref{eq:sfsurf}), with the uniform flow accounted for) may have some associated bound circulation, and it is desirable to adjust it by adding or subtracting a uniform value so that it has none. Equation (\ref{eq:circs0}) suggests that this circulation can be removed by setting $\szero$ to $\ipscalar{\onesspoint}{ \schurA^{-1}\sfsurfp}/\circzero\equiv\vsheetzeroop \sfsurfp/\circzero$ and then subtracting this value (multiplied by the uniform vector $\onesspoint$) from $\sfsurfp$. Overall, this process can be encapsulated in a {\em circulation removal} operator, $\projectC$, that acts upon a surface streamfunction, $\spoint{\sfbase} \in \spoints$,
\begin{equation}
\label{eq:circrem}
\projectC  \equiv \id - \frac{\onesspoint \vsheetzeroop}{\circzero}.
\end{equation}
It is easy to verify that $\ipscalar{\schurA^{-1}\onesspoint}{\projectC \spoint{\sfbase}} = \vsheetzeroop \projectC \spoint{\sfbase} = 0$ for any $\spoint{\sfbase} \in \spoints$, so that the circulation of any surface streamfunction acted upon by $\projectC$ is indeed zero. It is important to observe, also, that $\sfsurfp$ can be replaced by $\projectC \sfsurfp$ without affecting the nature of the no-penetration condition. We also note that the composite operator $\schurA^{-1}\projectC$ is symmetric, just as $\schurA^{-1}$ is, since
\begin{equation}
    \schurA^{-1}\projectC  = \schurA^{-1} - \vsheetzeroopbase \vsheetzeroop.
\end{equation}

\subsection{The Kutta condition}
\label{sec:kutta}

For surfaces that contain convex edges, the vortex sheet strength assumes a singular behavior in the vicinity of these edges, with a strength that depends on the interior angle of the edge: sharper edges have more singular behavior. In the discrete representation of the surface, edges are only approximately represented by the sudden disruptions of positions in clusters of adjacent points, without any information about the surface normals. The behavior in this discrete form is not quite singular, but the solution of (\ref{eq:vsheet}) nonetheless exhibits a large and rapid change of amplitude.

If we seek to eliminate this behavior, we must first have some means of exposing it. In fact, for any discretized surface, the essence of this nearly-singular behavior lies in the vector $\vsheetzero$, and all other bound vortex sheets associated with the same surface share the same nearly-singular behavior. Thus, we will use a multiplicative decomposition of the vortex sheet strength:
\begin{equation} \label{eq:decomposition}
\vsheet = \vsheetzero \had \svsheet,
\end{equation}
where $\had$ is the Hadamard product. This decomposed form isolates the singular behavior into $\vsheetzero$, and $\svsheet$ is a relatively smoother vector of surface point data. In the regularization operation on $\vsheet$, we can absorb $\vsheetzero$ into $\regds$, first noting that the Hadamard product can alternatively be written with the help of a diagonal matrix,
\begin{equation}
\vsheet = \vsheetzero \had \svsheet = \diagmat{\vsheetzero} \svsheet.
\end{equation}
Then, we can define a re-scaled regularization operator,
\begin{equation}
\regds\vsheet = \regds \diagmat{\vsheetzero} \svsheet = \regdssmooth \svsheet.
\end{equation}
The re-scaled operator $\regdssmooth = \regds \diagmat{\vsheetzero}$ can, in turn, be absorbed into the Schur complement, defining $\schurAsmooth = -\interp \lapgrid^{-1} \regdssmooth = \schurA \diagmat{\vsheetzero}$. A useful property of $\schurAsmooth$ is that it preserves uniform vectors:
\begin{equation}
\label{eq:Son1}
\schurAsmooth \onesspoint = \schurA \diagmat{\vsheetzero} \onesspoint = \schurA \vsheetzero = \onesspoint, \qquad \schurAsmooth^{-1}\onesspoint = \onesspoint.
\end{equation}

\begin{figure}
    \centering
    \includegraphics[page=1]{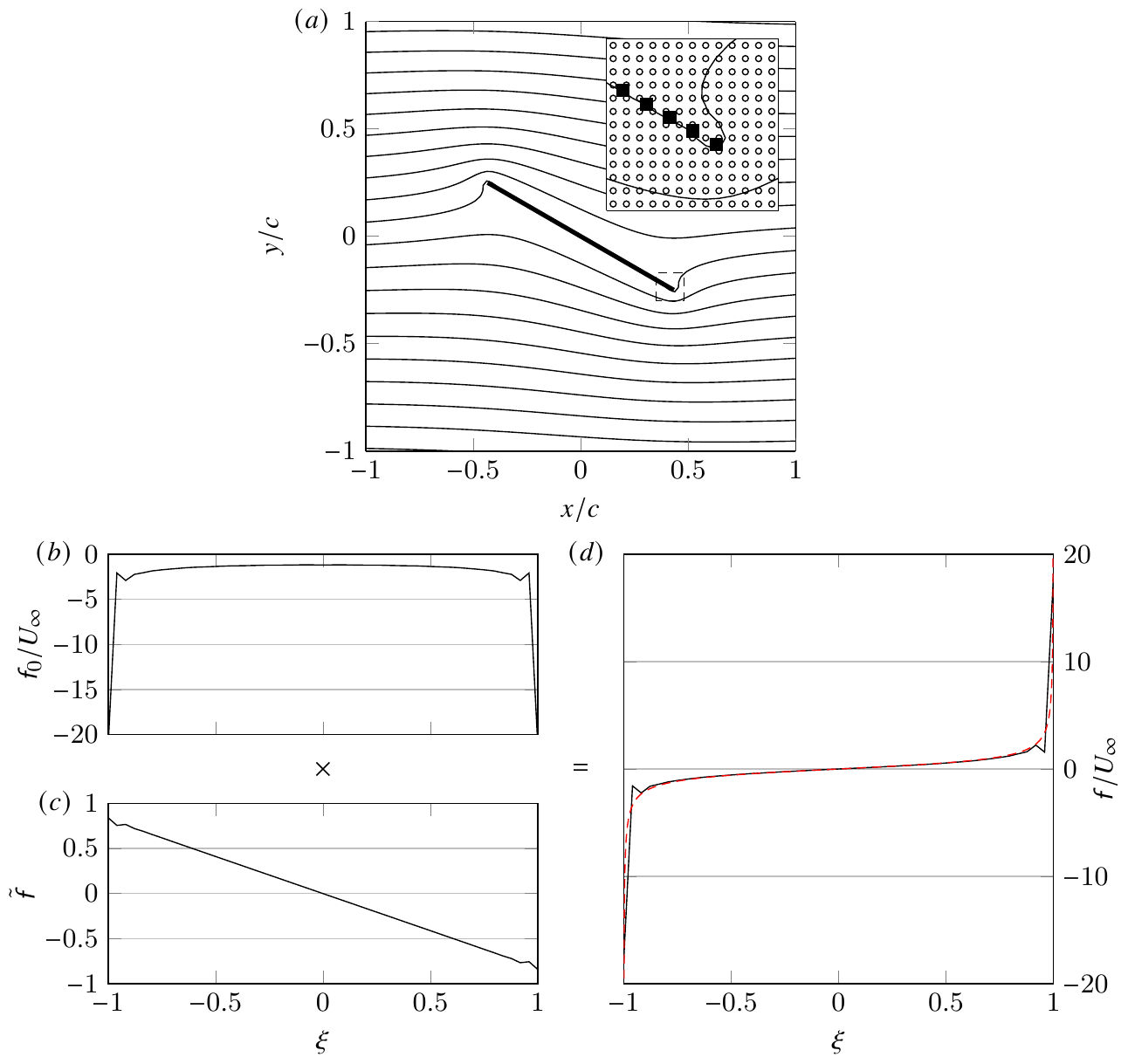}
    \caption{(\textit{a}) Contours (\full) of the discrete streamfunction for a flat plate of length $c$, at \SI{30}{\degree} in a uniform flow $U_\infty$ without enforcement of the Kutta condition. The inset figure shows a closeup of the nodes (\opencirc) and surface points (\fullsquare) in the boxed area (\dashed). The point-wise product of (\textit{b}) the discrete vortex sheet strength associated with a uniform, unit-strength streamfunction on the body, and (\textit{c}) the non-singular vector that results from re-scaling the regularization operator, composes (\textit{d}) the discrete vortex sheet strength. Overlaid is the exact continuous solution  (\textcolor{red}{\dashed}). The simulation is performed with $\dx/c=0.01$ and $\dS/\dx=2$.}
    \label{fig:nokutta}
\end{figure}

The decomposition of the vortex sheet strength is demonstrated in figure~\ref{fig:nokutta} for a flat plate in a uniform flow. We plot the vortex sheet strength against the scaled coordinate $\xi$, which varies along the plate from $-1$ at the leading edge to $1$ at the trailing edge. As expected, the vortex sheet strength $\vsheet$ shows large amplitude variations at the sharp leading and trailing edges, corresponding to the singularities in the distribution of the continuous vortex sheet strength. By use of decomposition (\ref{eq:decomposition}), these discrete singularities are retained in $\vsheetzero$ and we are left with a non-singular $\svsheet$, which varies almost linearly with $\xi$.

The Kutta condition corresponds to annihilating the nearly-singular behavior at a surface point. At such points, we will set the corresponding value of $\svsheet$ to zero. Suppose we wish to enforce the Kutta condition at an edge corresponding to surface point $k$. The condition is
\begin{equation}
\label{eq:kutta}
\unitspoint{k}^{T} \svsheet = 0.
\end{equation}


\subsubsection{Using the Kutta condition in a steady-state problem}

We will first take the steady-state approach to enforce the Kutta condition: allow the bound circulation to be set appropriately, with the implicit understanding that there is a starting vortex of equal and opposite circulation at infinity that preserves the Kelvin circulation theorem. The Lagrange multiplier for this constraint will not be $\circsurf$, but $\szero$, similar to (\ref{eq:circsystem}). We also use the circulation removal operator to adjust the imposed surface streamfunction:
\begin{equation}
\label{eq:kuttasystem}
\begin{bmatrix}
\lapgrid & \regdssmooth & 0 \\
\interp & 0 & \onesspoint \\
0 & \unitspoint{k}^{T}  & 0
\end{bmatrix}
\begin{pmatrix}
\sfgrid \\
\svsheet \\
\szero
\end{pmatrix} = 
\begin{pmatrix}
-\vortgrid \\
\projectC\sfsurfp  \\
0
\end{pmatrix}
\end{equation}

This block system, like the earlier one in (\ref{eq:basicblock}), has a saddle point form, and we can reduce it by the same block-LU decomposition to develop a solution algorithm. We will interpret it in the general form (\ref{eq:saddlept}), with the upper left $2\times 2$ block taking the role of $\basicmat$, the solution vector $\basicx$ and constraint force $\basicy$ set, respectively, to
\begin{equation}
\basicx = \begin{pmatrix} \sfgrid \\
\svsheet \end{pmatrix},\qquad  \basicy = \szero,
\end{equation}
the remaining operators set to
\begin{equation}
\basicBtwo = \begin{bmatrix} 0 & \unitspoint{k}^{T} \end{bmatrix}, \qquad \basicBoneT = \begin{bmatrix} 0 \\ \onesspoint \end{bmatrix},\qquad \basicC = 0,
\end{equation}
and the right-hand side vectors set to
\begin{equation}
\basicrone = \begin{pmatrix} -\vortgrid \\
\projectC\sfsurfp \end{pmatrix},\qquad  \basicrtwo = 0.
\end{equation}

We note that block $\basicmat$ has the original form of the system before the Kutta constraint (\ref{eq:basicblock}), though with the slight modification of a re-scaled regularization operator, and we already have the inverse of $\basicmat$ available from (\ref{eq:Ainverse}). The solution of this original system forms the  intermediate solution of the full system endowed with the Kutta condition:
\begin{equation}
\label{eq:intermedsoln}
\svsheet^{*}  = \schurAsmooth^{-1} \left(\projectC\sfsurfp + \interp\invlapgrid\vortgrid \right),\qquad \sfgrid^{*} = -\invlapgrid\left(\vortgrid + \regdssmooth \svsheet^{*}\right).
\end{equation}

Then, using the general procedure outlined in appendix \ref{app:saddlesystems}, the solution of the full system (\ref{eq:kuttasystem}) is easy to develop; its Schur complement is simply
\begin{equation}
\basicS = -1.
\end{equation}
Applying the general solution equations, and using the property (\ref{eq:Son1}) to simplify the resulting operators, it can be shown that the solution is
\begin{align}
\szero &= \unitspoint{k}^{T} \svsheet^{*} \label{eq:s0}\\
\begin{pmatrix} \sfgrid \\
\svsheet \end{pmatrix} &= \begin{pmatrix} \sfgrid^{*} \\
\svsheet^{*} \end{pmatrix} - \begin{bmatrix} -\invlapgrid\regdssmooth \onesspoint\\ \onesspoint\end{bmatrix}  \unitspoint{k}^{T} \svsheet^{*}.
\end{align}

The entire solution can be written more compactly as
\begin{align}
\svsheet &= \projectK{k} \schurAsmooth^{-1} \left( \projectC\sfsurfp + \interp\invlapgrid \vortgrid \right), \\
\sfgrid &= -\invlapgrid\left(\vortgrid  + \regdssmooth \svsheet \right),
\end{align}
where we have defined the {\em Kutta projection operator},
\begin{equation}
\projectK{k} \equiv \id -  \onesspoint \unitspoint{k}^{T},
\end{equation}
which acts upon the (non-singular part of the) bound vortex sheet vector, subtracting the value at point $k$ from every point, including at $k$ itself.

Note that the Lagrange multiplier for the Kutta condition takes the simple value given by equation (\ref{eq:s0}), revealing that the additional streamfunction on the surface is exactly the value of the intermediate bound vortex sheet at the Kutta point $k$.

\begin{figure}
    \centering
    \includegraphics[page=1]{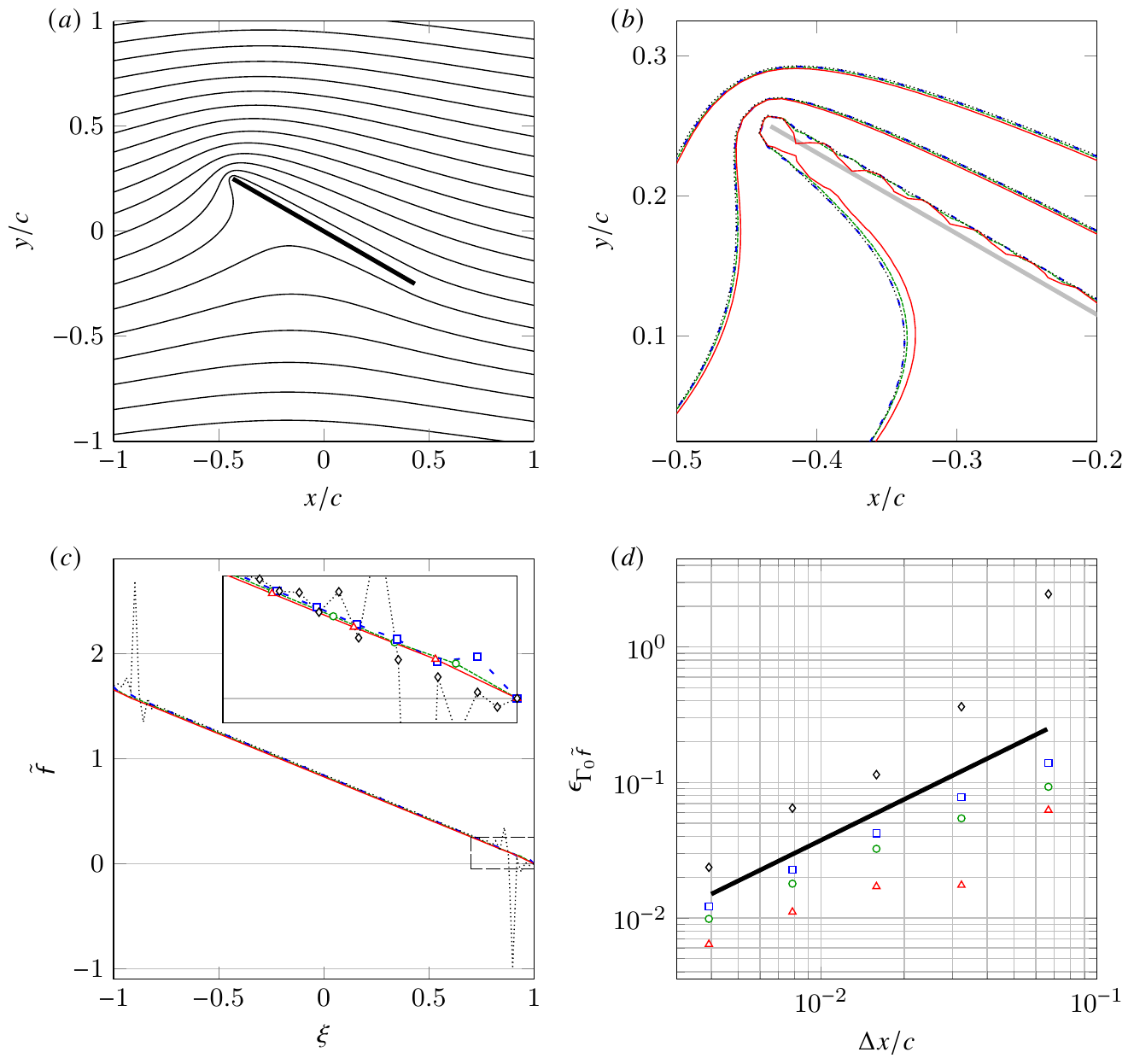}
    \caption{(\textit{a}) Contours ({\full}) of the steady, discrete streamfunction for a flat plate of length $c$ at \SI{30}{\degree} in a uniform flow with enforcement of the Kutta condition at the trailing edge. (\textit{b})
    Zoom on the leading edge showing the streamlines for three values of the streamfunction, (\textit{c}) the non-singular part of the associated discrete vortex sheet strength for $\dx/c=0.01$ with an inset enlarging the boxed area (\longbroken), and (\textit{d}) the variation of its error with grid spacing for different values for $\dS/\dx$: \dotted\, and  \opendiamond, $\dS/\dx=1$; \textcolor{blue}{\textbf{\broken}}\, and \textcolor{blue}{\opensquare}, $\dS/\dx=2$; \textcolor{green!60!black}{\dashed}\, and  \textcolor{green!60!black}{\opencirc}, $\dS/\dx=3$; \textcolor{red}{\full}\, and  \textcolor{red}{\opentriangle}, $\dS/\dx=4$. Overlaid is an error (\textbf{\full}) that scales as $\Delta x$.}
    \label{fig:steadykutta}
\end{figure}

The application of the Kutta condition to a steady-state problem is demonstrated in Figure~\ref{fig:steadykutta} on the flat plate problem that was introduced in the previous section. By constraining the trailing edge point of $\svsheet$, its whole distribution is shifted upward such that the trailing-edge value equals zero. The resulting streamfunction indicates that the flow then indeed leaves the trailing edge smoothly.

Figure~\ref{fig:steadykutta} also shows a spatial grid refinement analysis of the non-singular part of the vortex sheet strength. For this analysis, we can use a multiplicative decomposition for the continuous vortex sheet strength, $\gamma(\xi) = \gamma_0(\xi) \tilde{\gamma}(\xi)$, analogous to~(\ref{eq:decomposition}). The continuous counterpart of $\vsheetzero$, can be found as the bound vortex sheet strength of a circular cylinder with a point vortex of strength $\circzero$ at its center, conformally mapped to a flat plate. Due to the definition of $\vsheetzero$, the circulation $\circzero$ depends on the grid spacing. Therefore, if we define $\gamma_0$ as the continuous counterpart of $\vsheetzero/\circzero$, we obtain the grid-independent solution $\tilde{\gamma}(\xi) = - \pi c U_\infty \sin(\alpha) (1-\xi)$ as the continuous counterpart of $\circzero\svsheet$ for a flat plate with the Kutta condition enforced. We can then define the error for the non-singular part of the vortex sheet strength as $\epsilon_{\circzero\svsheet} = \Vert \circzero\svsheet - \tilde{\gamma}(\xi(\xsurf,\ysurf)\Vert_2 / \Vert \tilde{\gamma}(\xi(\xsurf,\ysurf)\Vert_2$. The refinement analysis verifies that the immersed-boundary projection method using the vortex sheet strength decomposition is approximately first-order accurate in $\dx$. Furthermore, the figure shows the streamlines near the plate, the vortex sheet strength, and the error for four different values of $\dS/\dx$. These confirm that a lower value increases the noise in the vortex sheet strength, but a higher value can lead to the streamlines penetrating the surface, as discussed previously.

\subsubsection{Using the Kutta condition to set a new vortex element}

In the previous section, we used the Kutta condition to set the bound circulation but did not explicitly create a new vortex element. This vortex element was assumed to lie at infinity so that its effect was negligible except insofar as it left equal but opposite circulation about the body.

In this section, we will create a new vortex element in the vicinity of the edge at which we are applying the Kutta condition. We will thus seek to establish the strength of this new element and to do so in such a manner that the overall circulation of the flow is conserved. Once the element is created, it will be allowed to advect with the local fluid velocity.

Let us assume that the new vortex element (which we label with the subscript 1) is introduced at some point in physical space, and that its immersion into the Cartesian grid is described by $\unitgridvort{1}$ and that its strength (i.e., its circulation) is $\vortcirc{1}$. Thus, the fluid vorticity after this new element's introduction can be written as
\begin{equation}
\vortgrid + \vortcirc{1}\unitgridvort{1}.
\end{equation}

The Kutta condition (\ref{eq:kutta}) is still to be enforced. We also seek to ensure that the total circulation is zero to satisfy Kelvin's circulation theorem. (We are assuming that the flow has started from rest.) Let us denote the circulation of the existing fluid vorticity $\vortgrid$ by 
\begin{equation}
\circw = \ipnodes{\onesgrid}{\vortgrid}.
\end{equation}
Then, the circulation constraint is
\begin{equation}
\ipscalar{\onesspoint}{\vsheet} + \vortcirc{1} + \circw = 0.
\end{equation}
The circulation of the bound vortex sheet $\vsheet$ can be re-written in terms of the non-singular part of the sheet as $\ipscalar{\onesspoint}{\vsheet} = \ipscalar{\vsheetzero}{\svsheet} = \vsheetzeroop \svsheet$.

With these two constraints, the overall saddle point system of equations is
\begin{equation}
\label{eq:kuttasystem2}
\begin{bmatrix}
\lapgrid & \regdssmooth & 0 & \unitgridvort{1} \\
\interp & 0 & \onesspoint & 0\\
0 & \unitspoint{k}^{T}  & 0 & 0 \\
0 & \vsheetzeroop & 0 & 1
\end{bmatrix}
\begin{pmatrix}
\sfgrid \\
\svsheet \\
\szero \\
\vortcirc{1} 
\end{pmatrix} = 
\begin{pmatrix}
-\vortgrid \\
\projectC\sfsurfp  \\
0 \\
-\circw
\end{pmatrix}.
\end{equation}
Again, the basic saddle-point matrix constitutes the upper left $2\times2$ block $\basicmat$ and the solution vector $\basicx$ is as before. The constraint force vector is
\begin{equation}
\basicy = \begin{pmatrix} \szero \\
\vortcirc{1}\end{pmatrix},
\end{equation}
and the remaining vectors and operators are now
\begin{equation}
\basicrtwo = \begin{pmatrix} 0 \\
-\circw \end{pmatrix},\qquad \basicBtwo = \begin{bmatrix} 0 & \unitspoint{k}^{T} \\ 0 & \vsheetzeroop \end{bmatrix}, \qquad \basicBoneT = \begin{bmatrix} 0 & \unitgridvort{1} \\ \onesspoint & 0\end{bmatrix},\qquad \basicC =-\begin{bmatrix}  0 & 0 \\ 0 & 1\end{bmatrix}.
\end{equation}

The solution algorithm follows, once again, from the equations in appendix \ref{app:saddlesystems}. After carrying out the block matrix multiplications, it can be shown that the Schur complement (\ref{eq:schur}) is the $2\times 2$ matrix
\begin{equation}
\basicS = \begin{bmatrix} -1 & \unitspoint{k}^{T} \unitsvsheet{1} \\ -\vsheetzeroop \onesspoint & 1+\vsheetzeroop \unitsvsheet{1}, \end{bmatrix}
\end{equation}
where, for convenience, we have defined
\begin{equation}
\unitsvsheet{1} = \schurAsmooth^{-1} \interp \invlapgrid \unitgridvort{1},
\end{equation}
which represents the (non-singular part of the) strength of the vortex sheet that ``reacts'' to the presence of a unit-strength vortex $\unitgridvort{1}$ immersed into the grid, canceling that vortex's induced velocity on the surface. The term $\vsheetzeroop \unitsvsheet{1}$ represents this sheet's bound circulation and $\unitspoint{k}^{T} \unitsvsheet{1}$ is its contribution to the Kutta condition at point $k$. The problem (\ref{eq:algor}) for the constraint forces $\szero$ and $\vortcirc{1}$ is then
\begin{equation}
\begin{bmatrix} -1 & \unitspoint{k}^{T} \unitsvsheet{1} \\ -\vsheetzeroop \onesspoint & 1+\vsheetzeroop \unitsvsheet{1}, \end{bmatrix} \begin{pmatrix} \szero \\
\vortcirc{1}\end{pmatrix} = \begin{pmatrix} -\unitspoint{k}^{T} \svsheet^{*} \\
-\circw - \vsheetzeroop  \svsheet^{*} \end{pmatrix},
\end{equation}
where the intermediate solution $\svsheet^{*}$ is available from (\ref{eq:intermedsoln}).

The determinant of this Schur complement matrix is $-1 - \vsheetzeroop \projectK{k} \unitsvsheet{1}$, which represents the negative of the circulation of the unit vortex and its associated vortex sheet, after the Kutta condition has been enforced on this sheet. It is straightforward then to calculate the strength of the new vortex $\vortcirc{1}$ and the additional uniform surface streamfunction, $\szero$:
\begin{equation}
\vortcirc{1} = \frac{-\circw - \vsheetzeroop \projectK{k} \svsheet^{*}}{1 + \vsheetzeroop \projectK{k} \unitsvsheet{1}},\qquad \szero = \unitspoint{k}^{T} \left(\svsheet^{*} + \vortcirc{1}\unitsvsheet{1}  \right).
\end{equation}
From these, we can then obtain the vortex sheet strength and the fluid streamfunction,
\begin{equation}
\svsheet = \projectK{k} \left( \svsheet^{*} + \vortcirc{1} \unitsvsheet{1} \right),\qquad \sfgrid = -\invlapgrid\left(\vortgrid +\vortcirc{1}\unitgridvort{1} + \regdssmooth \svsheet\right).
\end{equation}
The intermediate solution, which corresponds to the flow associated with existing vorticity, is corrected here with the new vortex $1$ to enforce the Kutta condition at the point $k$.

\begin{figure}
    \centering
    \includegraphics[page=1]{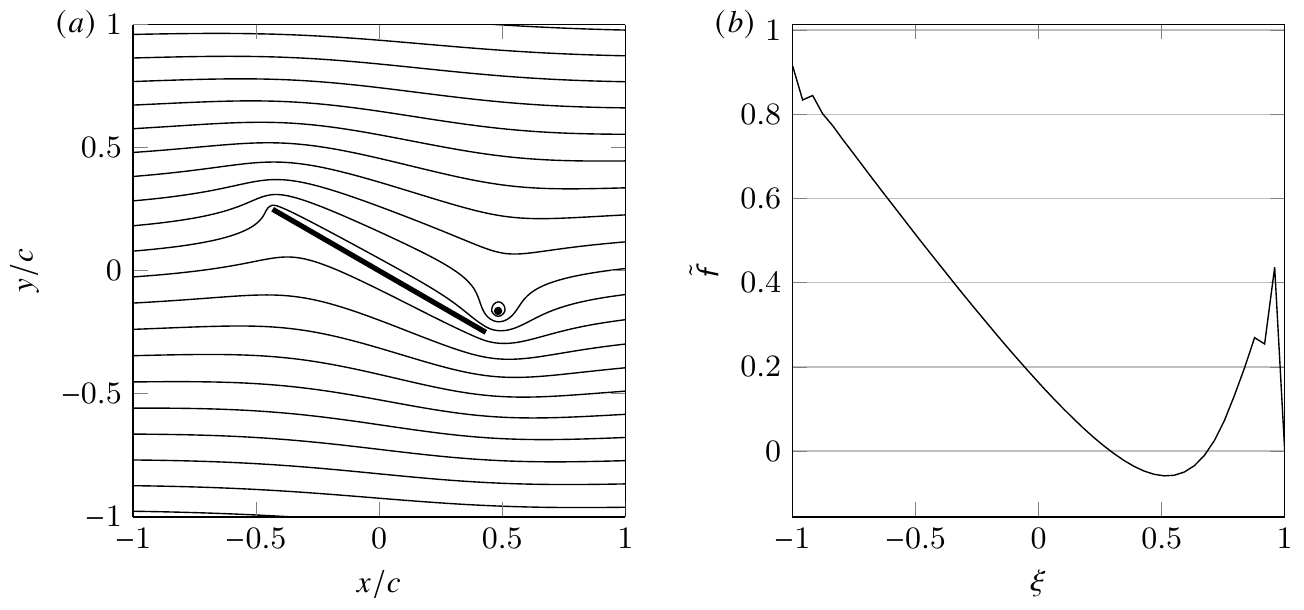}
    \caption{(\textit{a}) Contours (\full) of the unsteady, discrete streamfunction for a flat plate of length $c$ at \SI{30}{\degree} in a uniform flow with release of vorticity into a point vortex (\fullcirc) for enforcement of the Kutta condition at the trailing edge. (\textit{b}) The non-singular part of the associated discrete vortex sheet strength. The simulation is performed with $\dx/c=0.01$ and $\dS/\dx=2$.}
    \label{fig:unsteadykutta}
\end{figure}

\begin{figure}
    \centering
    \includegraphics[page=1]{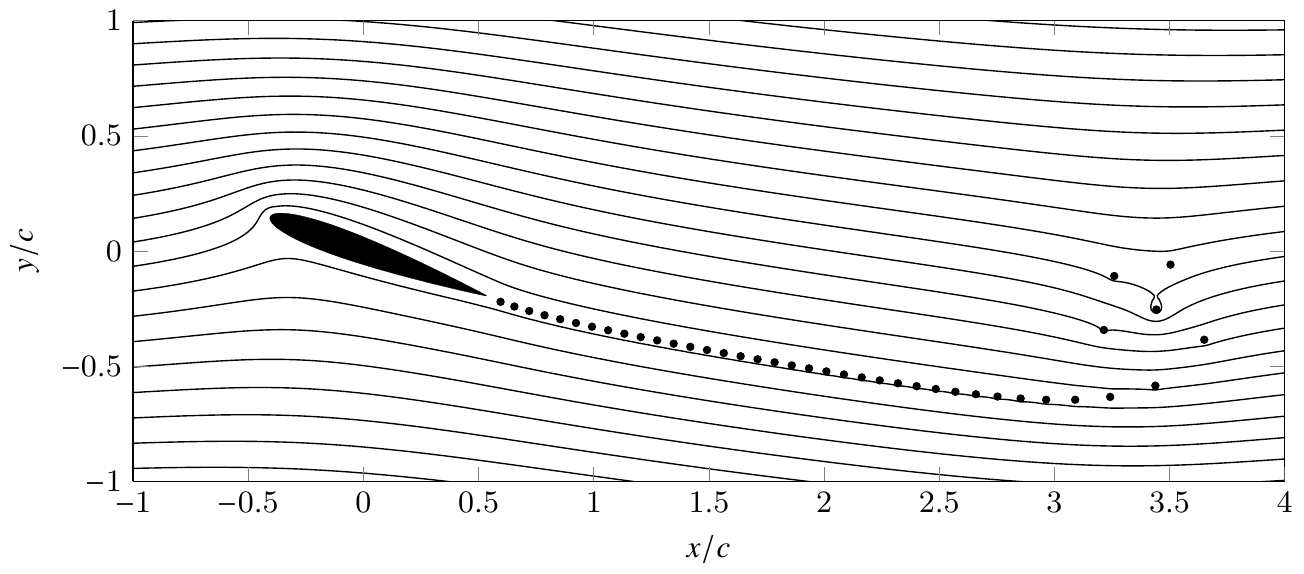}
    \caption{Contours (\full) of the unsteady, discrete streamfunction for a NACA0012 airfoil with chord length $c$ at \SI{20}{\degree}, 3 convective times after impulsively starting a uniform flow $U_{\infty}$. Vorticity is released into a stream of point vortices (\fullcirc) for enforcement of the Kutta condition at the trailing edge. The simulation is performed with $\dx/c=0.01$, $\dS/\dx=1.5$, and $\Delta t U_{\infty}/c=0.075$.}
    \label{fig:thickairfoil}
\end{figure}

We now first demonstrate the enforcement of the Kutta condition in unsteady flow on the flat plate problem with a point vortex near the trailing edge to enforce the Kutta condition at that edge. We position the point vortex at a distance $10 \Delta t  U_{\infty}$ from the edge in the direction of the free stream, perpendicular to the plate. Figure~\ref{fig:unsteadykutta} demonstrates that, because of the proximity of the point vortex to the flat plate, $\svsheet$ exhibits a quick variation at the surface points that lie closest to the point vortex. The value at the trailing edge point itself is still constrained to zero and the flow again leaves the edge smoothly. This situation corresponds to the flow right after impulsively starting a uniform flow around a flat plate and the point vortex now represents the starting vortex. The second demonstration of the method (figure~\ref{fig:thickairfoil}) is the repeated enforcement of the Kutta condition on a NACA0012 airfoil while advancing the positions of the point vortices in time after impulsively starting the flow. This simulation, and all following simulations, use forward Euler time-stepping, unless noted otherwise, and new point vortices are inserted at one-third of the way from the edge to the last released vortex from that edge. The streamlines in the figure show that a strong initial vortex and weaker subsequent vortices were created and convected downstream. In turn, at each time step, the airfoil obtained a circulation that enforces the Kutta condition.

\subsubsection{Applying more than one Kutta condition on a body}
\label{sec:kutta2}

Suppose we wish to enforce the Kutta condition at two edges of the body---at points $k_{1}$ and $k_{2}$---instead of one. Each such point has a constraint,
\begin{equation}
\unitspoint{k_{j}}^{T} \svsheet = 0,\quad j = 1,2.
\end{equation}
For two such constraints, we need two Lagrange multipliers: the strengths of two new vortices, $\vortcirc{1}$ and $\vortcirc{2}$, immersed into the grid with $\unitgridvort{1}$ and $\unitgridvort{2}$, respectively; and we still need the Lagrange multiplier $\szero$ to ensure that Kelvin's circulation theorem is also enforced. The system in the previous section is thus easily generalized to the following:
\begin{equation}
\label{eq:kuttasystem3}
\begin{bmatrix}
\lapgrid & \regdssmooth & 0 & \unitgridvort{1} & \unitgridvort{2} \\
\interp & 0 & \onesspoint & 0 & 0\\
0 & \unitspoint{k_{1}}^{T}  & 0 & 0 & 0 \\
0 & \unitspoint{k_{2}}^{T}  & 0 & 0 & 0 \\
0 & \vsheetzeroop & 0 & 1 & 1
\end{bmatrix}
\begin{pmatrix}
\sfgrid \\
\svsheet \\
\szero \\
\vortcirc{1} \\
\vortcirc{2} 
\end{pmatrix} = 
\begin{pmatrix}
-\vortgrid \\
\projectC\sfsurfp  \\
0 \\
0 \\
-\circw
\end{pmatrix}.
\end{equation}

The system is reduced in the same manner as before, with the same intermediate solution obtained from the basic system (\ref{eq:basicblock}). Now, the Schur complement problem for the constraint forces takes the form
\begin{equation}
\label{eq:schurkutta2}
\begin{bmatrix} -1 & \unitspoint{k_{1}}^{T} \unitsvsheet{1} & \unitspoint{k_{1}}^{T} \unitsvsheet{2} \\-1 & \unitspoint{k_{2}}^{T} \unitsvsheet{1} & \unitspoint{k_{2}}^{T} \unitsvsheet{2} \\  -\vsheetzeroop \onesspoint & 1+\vsheetzeroop \unitsvsheet{1} & 1+\vsheetzeroop \unitsvsheet{2}, \end{bmatrix} \begin{pmatrix} \szero \\
\vortcirc{1}\\ \vortcirc{2}\end{pmatrix} = \begin{pmatrix} -\unitspoint{k_{1}}^{T} \svsheet^{*} \\ -\unitspoint{k_{2}}^{T} \svsheet^{*} \\
-\circw - \vsheetzeroop  \svsheet^{*} \end{pmatrix},
\end{equation}
where we have now defined bound vortex sheets associated with each of the two new vortices (with unit strengths): 
\begin{equation}
\unitsvsheet{j} = \schurAsmooth^{-1} \interp \invlapgrid \unitgridvort{j},
\end{equation}
for $j = 1,2$. It is interesting to note that, if we take the difference between the two Kutta constraints, we obtain
\begin{equation}
\label{eq:kuttadiff}
    \left(\unitspoint{k_{1}}^T- \unitspoint{k_{2}}^T\right)\left( \svsheet^{*} + \vortcirc{1} \unitsvsheet{1} + \vortcirc{2} \unitsvsheet{2}\right) = 0.
\end{equation}

It can be shown that this Schur complement problem can be split into
\begin{equation}
\label{eq:twovortex}
\begin{bmatrix}
1 + \vsheetzeroop\projectK{k_{1}}\unitsvsheet{1} & 1 + \vsheetzeroop\projectK{k_{1}}\unitsvsheet{2} \\  1 + \vsheetzeroop\projectK{k_{2}}\unitsvsheet{1} & 1 + \vsheetzeroop\projectK{k_{2}}\unitsvsheet{2} 
\end{bmatrix} \begin{pmatrix} \vortcirc{1}\\ \vortcirc{2} \end{pmatrix} = -\begin{pmatrix} \circw + \vsheetzeroop \projectK{k_{1}}\svsheet^{*}\\ \circw + \vsheetzeroop \projectK{k_{2}}\svsheet^{*}\end{pmatrix}
\end{equation}
and 
\begin{equation}
\szero = \frac{1}{2} \left(\unitspoint{k_{1}}^{T}+\unitspoint{k_{2}}^{T}\right)\left(\svsheet^{*} + \vortcirc{1} \unitsvsheet{1} + \vortcirc{2} \unitsvsheet{2} \right).
\end{equation}
The latter equation, when combined with (\ref{eq:kuttadiff}), reveals that the value of the vortex sheet strength $\svsheet^{*} + \vortcirc{1} \unitsvsheet{1} + \vortcirc{2} \unitsvsheet{2}$ is the same at both Kutta points and equal to $\szero$.

Equation (\ref{eq:twovortex}) can be solved easily for the strengths of the two new point vortices. Then, the solution for the vortex sheet strength and streamfunction are
\begin{equation}
\svsheet = \frac{1}{2}\left( \projectK{k_{1}} + \projectK{k_{2}} \right)\left(\svsheet^{*} + \vortcirc{1} \unitsvsheet{1} + \vortcirc{2} \unitsvsheet{2} \right),\qquad \sfgrid = -\invlapgrid\left(\vortgrid +\vortcirc{1}\unitgridvort{1} + \vortcirc{2}\unitgridvort{2} + \regdssmooth \svsheet\right)
\end{equation}

\begin{figure}
    \centering
    \includegraphics[page=1]{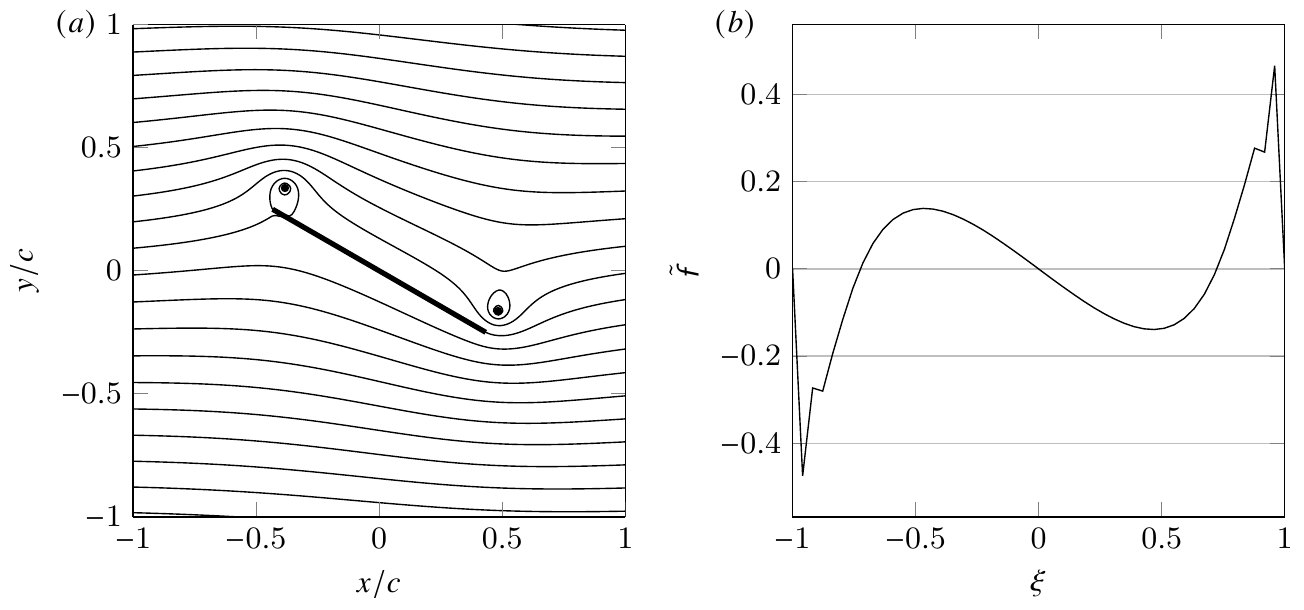}
    \caption{(\textit{a}) Contours (\full) of the unsteady, discrete streamfunction for a flat plate of length $c$ at \SI{30}{\degree} in a uniform flow with release of vorticity into two point vortices (\fullcirc) for enforcement of the Kutta condition at both edges. (\textit{b}) The non-singular part of the associated discrete vortex sheet strength. The simulation is performed with $\dx/c=0.01$ and $\dS/\dx=2$.}
    \label{fig:unsteadymultikutta}
\end{figure}

We now apply this method in figure~\ref{fig:unsteadymultikutta} to enforce the Kutta condition at the leading and trailing edge of our flat plate problem. We position a point vortex close to each edge and observe again that $\svsheet$ shows strong variation at the surface points closest to the two point vortices. The contours of the streamfunction indicate that the flow indeed leaves the edges smoothly. Like the previous case, this solution corresponds to the flow right after impulsively starting a uniform flow around a flat plate, but unlike the previous case, the flow now separates at the leading edge.

It should be observed that these solutions are posed in a manner easily extensible to an arbitrary number of edges.

\subsection{Generalized edge condition}
\label{sec:generalized}

In the previous section, we demonstrated the means of annihilating the (nearly) singular behavior at edges on a discretized surface. In some cases, our desire is not to annihilate this behavior, but simply to keep it within some bounds. In the analytical treatment of potential flow problems, this objective is served by placing an inequality constraint on the {\em edge suction parameter} \citep{ramesh14,darakjde2018,inviscidbook}. That parameter is proportional to the coefficient on the bound vortex sheet strength's singularity \citep{inviscidbook}, so in this discrete setting, in which we have extracted the singular part of $\vsheet$ in the form of $\vsheetzero$, we expect the suction parameter to be related to the value of $\svsheet$ at the edge. In fact, by simple comparison, it can be shown that
\begin{equation}
    \unitspoint{k}^T\svsheet = -\frac{2\pi c}{\circzero} \sucpar{k}
\end{equation}
for a flat plate of length $c$, where $\sucpar{k}$ is the suction parameter at the edge corresponding to point $k$.

Let $\sucparmin{k}$ and $\sucparmax{k}$ denote the minimum and maximum tolerable values of $\sucpar{k}$ at edge $k$. We then seek to confine the suction parameter to the range $\sucparmin{k} \leq \sucpar{k} \leq \sucparmax{k}$. This generalized edge constraint is placed on the suction parameter of the intermediate sheet $\svsheet^*$. To avoid confusion, we will redefine the bounds based on this non-singular part of the vortex sheet rather than $\sucpar{k}$ itself; for this, we define $\fmin{k} = -2\pi c \sucparmax{k}/\circzero$ and $\fmax{k} = -2\pi c \sucparmin{k}/\circzero$ if $\circzero$ is positive or $\fmax{k} = -2\pi c \sucparmax{k}/\circzero$ and $\fmin{k} = -2\pi c \sucparmin{k}/\circzero$ if $\circzero$ is negative. Thus, we inspect whether the value $\unitspoint{k}^T\svsheet^*$ lies in the range
\begin{equation}
    \fmin{k} \leq \unitspoint{k}^T\svsheet^* \leq \fmax{k}.
\end{equation}
If $\unitspoint{k}^T\svsheet^*$ lies within this range, then no new vortex is created near the edge (or equivalently, a new vortex of zero strength is created); if $\unitspoint{k}^T\svsheet^* > \fmax{k}$, then we create a new vortex so that $\unitspoint{k}^T\svsheet = \fmax{k}$; and if $\unitspoint{k}^T\svsheet^* < \fmin{k}$, then we do the same, but now so that $\unitspoint{k}^T\svsheet = \fmin{k}$. Note that the Kutta condition simply corresponds to setting $\fmin{k} = \fmax{k} = 0$.

We can easily accommodate these constraints into our solution approach for enforcing the Kutta condition from the previous section: in the case of two edges, by modifying the right-hand side vector of (\ref{eq:twovortex}) (if the edge suction lies outside of its bounds) or setting the vortex strength corresponding to that edge to zero. For example, suppose that $\unitspoint{k_1}^{T}\svsheet^{*} < \fmin{k_1}$ and $\unitspoint{k_2}^{T}\svsheet^{*} > \fmax{k_2}$; then we solve the system
\begin{equation}
\label{eq:twovortex-gen}
\begin{bmatrix}
1 + \vsheetzeroop\projectK{k_{1}}\unitsvsheet{1} & 1 + \vsheetzeroop\projectK{k_{1}}\unitsvsheet{2} \\  1 + \vsheetzeroop\projectK{k_{2}}\unitsvsheet{1} & 1 + \vsheetzeroop\projectK{k_{2}}\unitsvsheet{2} 
\end{bmatrix} \begin{pmatrix} \vortcirc{1}\\ \vortcirc{2} \end{pmatrix} = -\begin{pmatrix} \circw + \vsheetzeroop \projectK{k_{1}}\svsheet^{*}+\circzero\fmin{k_1}\\ \circw + \vsheetzeroop \projectK{k_{2}}\svsheet^{*}+\circzero\fmax{k_2}\end{pmatrix}.
\end{equation}
But if, say, $\fmin{k_2} \leq \unitspoint{k_2}^{T}\svsheet^{*} \leq \fmax{k_2}$, then we set $\vortcirc{2}=0$ and this system reduces to
\begin{equation}
    \vortcirc{1} = -\frac{\circw + \vsheetzeroop \projectK{k_{1}}\svsheet^{*}+\circzero\fmin{k_1}}{1 + \vsheetzeroop\projectK{k_{1}}\unitsvsheet{1}}.
\end{equation}

\begin{figure}
    \centering
    \includegraphics[page=1]{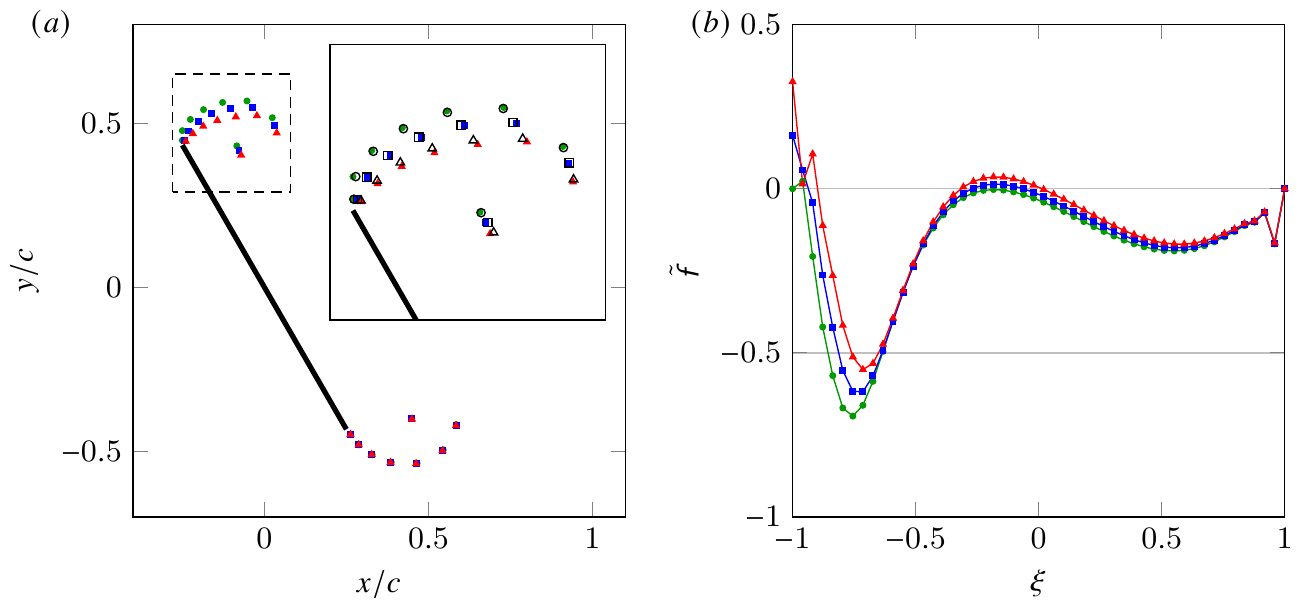}
    \caption{Effect of increasing $\sucparmax{\mathrm{LE}}/U_\infty$ from $0$ (\textcolor{green!60!black}{\fullcirc}) to $0.05$ (\textcolor{blue}{\fullsquare}) and $0.1$ (\textcolor{red}{\fulltriangle}) on (\textit{a}) the positions of shedded point vortices and (\textit{b}) the non-singular part of the associated discrete vortex sheet strength for a flat plate of length $c$ at \SI{60}{\degree}, $0.15$ convective times after impulsively starting a uniform flow $U_\infty$. The inset enlarges the boxed area (\longbroken) and overlays the positions of the vortices (\opencirc,\,\opensquare, and \opentriangle) obtained using the Biot-Savart method of~\citet{darakjde2018}. At the trailing edge, the Kutta condition is enforced. The simulation is performed with $\dx/c=0.01$, $\dS/\dx=2$, and $\Delta t U_{\infty}/c=0.025$.}
    \label{fig:lesp}
\end{figure}

The effect of applying these generalized edge conditions to the leading edge of a flat plate is shown in figure~\ref{fig:lesp} for the first instants after impulsively starting a uniform flow. The positions of the point vortices emanating from the leading edge in the figure indicate that as $\sucparmax{k}$ increases, the stream of point vortices is swept back from the edge. At the trailing edge, the Kutta condition is enforced in each case and the positions of the point vortices overlap, as they are not yet influenced by the different situations at the leading edge in these first instants.

\subsection{Pressure, force, and the added mass}
\label{sec:force}

In this section, we present the means for calculating pressure and force (and moment) in the grid-based treatment.

\subsubsection{Pressure}

Here we devise a means of computing the pressure, both throughout the flow field, $\pressgrid \in \centers$, as well as on the surface of a body. As we will show below, our immersed boundary treatment naturally gives rise to the jump in pressure across this surface, $\dpsurf = \psurf^+ - \psurf^- \in \spoints$ (with $+$ in the direction of the surface normal, $\normvecc{} \in \vecpoints$). Thus, to distinguish the pressures on either side of the surface from one another, we use the fact that the interpolation of $\pressgrid$ onto the surface produces the average of surface values, $\interpc \pressgrid = (\psurf^+ + \psurf^-)/2 \in \surface$, where $\interpc$ interpolates data from cell centers to the surface. It is thus easy to see that the pressure on either side is
\begin{equation}
    \psurf^{\pm} = \interpc \pressgrid \pm \tfrac{1}{2} \dpsurf.
\end{equation}

Thus, we seek $\dpsurf$ and $\pressgrid$. It should not be a surprise that our starting point for these quantities is the Euler equations. However, our approach exploits the fact that the methodology we have presented thus far already solves the Euler equations in the fluid---in vorticity form, via transport of vortex elements---and provides us with the instantaneous velocity field $\velgrid$ and strength of the bound vortex sheet $\vsheet$ on any bodies. This approach, which satisfies the incompressibility constraint by expressing velocity in the null space of the divergence operator (i.e., as curl of a streamfunction), obviates the need for computing pressure, the Lagrange multiplier for this constraint. However, now that we seek pressure, we use the Euler equations in their velocity form to provide it.

It is important to note that the immersed boundary treatment enriches the Euler equations' standard form with surface terms \citep{eldredge2021ilm}. Written in their spatially-discrete form, these immersed boundary Euler equations are
\begin{align}
\label{eq:euler}
    \rho \frac{\mathrm{d}\velgrid}{\mathrm{d}t} + \rho \left(\vortgrid + \regds \vsheet \right) \times \velgrid &= - \gradgrid \left(\pressgrid + \tfrac{1}{2}\rho |\velgrid|^2 \right) + \nonumber\\ & \regdsf \left[\normvecc{} \had \dpsurf + \rho \Delta \velgrid \left(\interpf\velgrid - \velsurf \right)\cdot\normvecc{}\right],
\end{align}
 where $\rho$ is the fluid density and $\interpf$ interpolates grid data from $\faces$ to the space of vector-valued surface data $\vecpoints$. Note the appearance of a few terms that are not typically seen in the Euler equations. First, we note the final bracketed pair of surface terms on the right-hand side, containing the jumps in pressure and velocity ($\Delta \velgrid$) across the surface, as well as the difference between the normal components of the fluid velocity (interpolated onto the surface) and the surface velocity itself, $\velsurf\in \vecpoints$. (This difference is zero by virtue of the no-flow-through condition, but we keep it here since it combines with other terms in later manipulations.) These surface terms are immersed into the grid by the operator $\regdsf$. In fact, if we had chosen to solve the Euler equations in velocity---instead of by streamfunction-vorticity---form, then we would have used the pressure jump in this term as a Lagrange multiplier for enforcing the no-flow-through condition on $\velgrid$. The second new term is the bound vortex sheet strength, appearing alongside the fluid vorticity on the left-hand side. This term emerges because the curl of the velocity field in the immersed boundary method generates both of these: $\curlgrid^T \velgrid = \vortgrid + \regds \vsheet$, where $\curlgrid^T: \faces \mapsto \nodes$. (Our notation for this operator is consistent with that of previous authors, such as \cite{Colonius2008}.)
 
These quantities, $\dpsurf$ and $\pressgrid$, can be solved for simultaneously from the Schur complement system that arises from solving the Euler equations (\ref{eq:euler}) and the associated constraints of divergence-free velocity and no-flow-through condition. Indeed, the approach we outline here is the natural outcome of that system. However, rather than present a detailed derivation, we present the equations with an intuitive explanation.

First, we develop an equation for $\dpsurf$ by taking the discrete curl $\curlgrid^T$ of (\ref{eq:euler}) to eliminate the gradient term---since $\curlgrid^T \gradgrid$ is identically zero---and obtain a vorticity form of the immersed-boundary Euler equations. With some manipulation to account for the motion of the surface (embodied in the time variation of $\regds$), these equations can be written as
\begin{equation}
    \rho \left( \frac{\mathrm{d}\vortgrid}{\mathrm{d}t} - \curlgrid^T (\velgrid\times \vortgrid)\right) + \rho \regds \frac{\mathrm{d}\vsheet}{\mathrm{d}t}  = \curlgrid^T \regdsf \normvecc{} \had\left[ \dpsurf + \rho \left(\interpf\velgrid-\velsurf\right)\cdot(\vsheet\times \normvecc{}) \right],
    \label{eq:surfaceeuler}
\end{equation}
where $\mathrm{d}\vsheet/\mathrm{d}t$ represents the time derivative of each element of $\vsheet$ while following a point moving with velocity $\velsurf$. We assume that each immersed point moves with this local surface velocity, so this is simply the time derivative of the vector $\vsheet$. We have written the equation in this intermediate form on purpose in order to make a few key points. First, it is important to note that we have already satisfied the vorticity equation in the fluid---the first terms in parentheses on the left-hand side---by advecting the point vortices. Thus, we can set these terms to zero, leaving only those terms associated with the surface. Aside from $\dpsurf$, these remaining terms involve only known quantities, and we could solve them in the current form for $\dpsurf$ using similar techniques to the ones we will describe below. However, we will first write the equation in a more familiar form, and define some helpful operators and quantities to enable this.


In equation (\ref{eq:surfaceeuler}) we see a composite of the curl of the regularization operator; let us write this more compactly as $\curlgridsurf^T: \spoints \mapsto \nodes$,  
\begin{equation}
    \curlgridsurf^T \spoint{\sigma}  = \curlgrid^T \regdsf \left(\normvecc{} \had \spoint{\sigma} \right),
\end{equation}
for some surface scalar data, $\spoint{\sigma} \in \spoints$. We refer to this as a {\em surface curl} operator. Its transpose, $\curlgridsurf: \nodes \mapsto \spoints$, also arises in what follows, and can also be described as a surface curl operator. It is defined as
\begin{equation}
    \curlgridsurf \sfgrid = \normvecc{}\cdot \interpf \curlgrid \sfgrid,
\end{equation}
for $\sfgrid \in \nodes$. The operator $\curlgridsurf$ obtains the normal component of velocity on an immersed surface for a given streamfunction $\sfgrid$. Before we explain the role of its transpose, it is useful to remember that any potential flow generated by (or about) an impenetrable surface can be equivalently described by either a distribution of vortices (a vortex sheet, with strength $\vsheet$) or a distribution of dipoles (a double layer) on the surface. In the latter case, the strength of the double layer is given by the negative of the jump in scalar potential across the surface, $-\Delta \spoint{\phiup} \in \spoints$. In fact, the two distributions' strengths can be related to each other, either by using Stokes' theorem or by the properties of the generalized functions that underpin the immersed boundary method \citep{eldredge2021ilm}, leading to
\begin{equation}
\label{eq:gamphi}
    \regds \vsheet = - \curlgridsurf^T \Delta \spoint{\phiup}.
\end{equation}
Thus, $\curlgridsurf^T$ produces the equivalent bound vorticity distribution (immersed into the grid) associated with a given jump in scalar potential on the surface. To calculate $\Delta \spoint{\phiup}$ in terms of $\vsheet$, we apply $\curlgridsurf \invlapgrid$ to both sides of (\ref{eq:gamphi}) to equate the normal velocity induced on the surface by each distribution. The composite operator $-\curlgridsurf\invlapgrid \curlgridsurf^T$ is positive semi-definite, and its null space can be shown to consist only of uniform values on the surface (i.e., the null space has an equivalent bound vorticity equal to zero). Thus, the jump in scalar potential associated with a vortex sheet of strength $\vsheet$ is
\begin{equation}
    \Delta \spoint{\phiup} = -\left(\curlgridsurf\invlapgrid \curlgridsurf^T \right)^{-1} \curlgridsurf\invlapgrid \regds \vsheet,
\end{equation}
to which we can add any constant value without affecting the result.

Now, armed with this insight, we can return to equation (\ref{eq:surfaceeuler}), and rewrite the surface terms on the left-hand side in terms of $\Delta \spoint{\phiup}$:
\begin{equation}
\label{eq:dphi}
    \curlgridsurf^T \left[ \dpsurf + \rho \left(\interpf\velgrid-\velsurf\right)\cdot(\vsheet\times \normvecc{}) + \rho \frac{\mathrm{d} \Delta \spoint{\phiup}}{\mathrm{d}t}\right] = 0,
\end{equation}
where $\mathrm{d} \Delta \spoint{\phiup}/\mathrm{d}t$ denotes the time derivative of $\Delta \spoint{\phiup}$ associated with a particular immersed point, assumed to be moving with velocity $\velsurf$. This equation implies that the expression in brackets must be equal to a uniform value, which we can take to be zero without loss of generality. We can immediately write an immersed boundary Bernoulli equation,
\begin{equation}
\label{eq:surfbernoulli}
    \dpsurf  + \rho \left(\interpf\velgrid-\velsurf\right)\cdot(\vsheet\times \normvecc{}) + \rho \frac{\mathrm{d} \Delta \spoint{\phiup}}{\mathrm{d}t} = 0.
\end{equation}
This equation is the discrete equivalent of a continuous version that appears in previous works, e.g. \cite{Jones2003,inviscidbook}. At each time step, we use (\ref{eq:dphi}) to compute the instantaneous jump in scalar potential associated with the vortex sheet $\vsheet$, and then use (\ref{eq:surfbernoulli}) to find $\dpsurf$.

Now we can substitute $\dpsurf$ from equation (\ref{eq:surfbernoulli}) into the Euler equations (\ref{eq:euler}) and operate on these equations with the discrete divergence operator, so that $\lapgrid = \divgrid\gradgrid$ acts on $\pressgrid$. We solve the resulting equation, obtaining
\begin{equation}
\label{eq:pgrid}
    \pressgrid = -\frac{1}{2} \rho |\velgrid|^2 - \rho \invlapgrid \left[ \divgrid (\vortgrid \times \velgrid) + \regdsf\left( \normvecc{} \had \frac{\mathrm{d}\Delta \spoint{\phiup}}{\mathrm{d}t} - \velsurf\times\vsheet \right) \right].
\end{equation}
It should be noted that any uniform value can be arbitrarily added to this expression. Also, we note in passing that the final set of terms (with the inverse Laplacian acting on the term in brackets) is equivalent to the time derivative of the scalar potential field, rendering the overall equation equivalent to a Bernoulli equation on the grid. However, there is no particular advantage in writing the equation in that form. In the current form, the first term in brackets (the divergence of the Lamb vector) represents the direct force exerted on the fluid by the fluid vorticity. The remaining terms in brackets collectively constitute the effects of surface motion and of the surface's modification of the flow induced by fluid elements (e.g., vorticity, free stream).

\begin{figure}
    \centering
    \includegraphics[page=1]{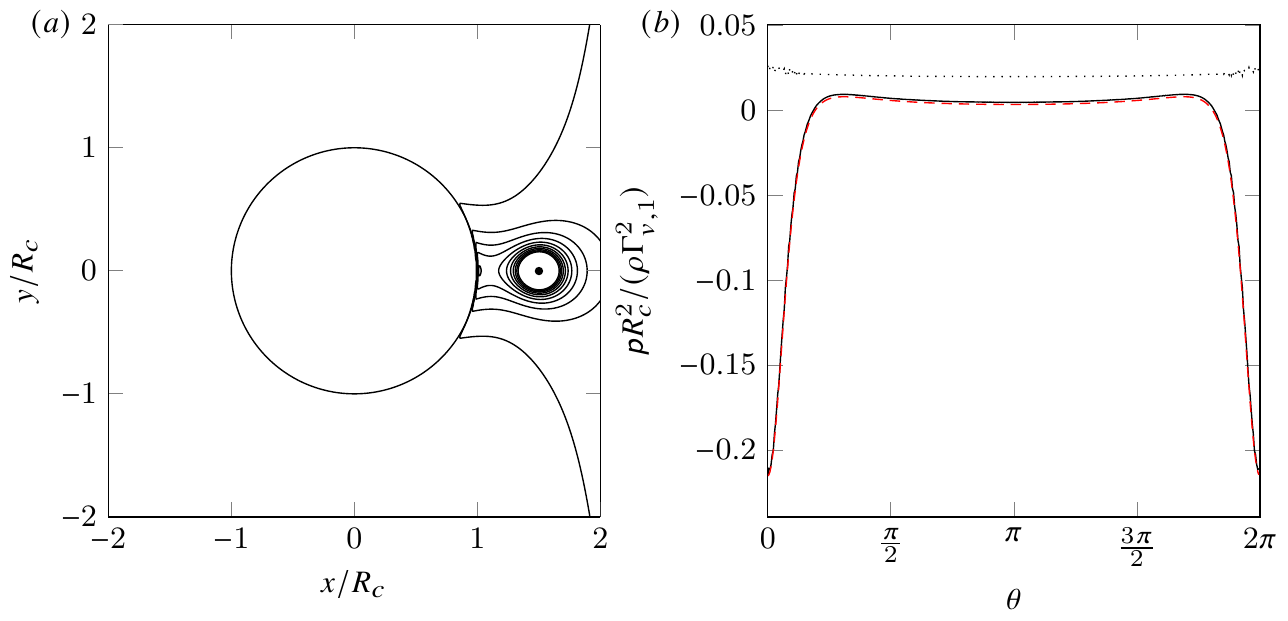}
    \caption{(\textit{a}) Contours (\full) of the discrete pressure for a point vortex (\fullcirc) with strength $\vortexstrengthq{1}$ at $(R_v,0)$ near a cylinder consisting of $\numpts$ points with radius $R_c$ and a bound circulation $-\vortexstrengthq{1}$. (\textit{b}) The scaled discrete pressure at the exterior (\full) and interior (\dotted) of the cylinder. Overlaid is the exact continuous solution  (\textcolor{red}{\dashed}) for the exterior pressure. The simulation is performed with $R_v/R_c = 3/2$, $\dx/R_c=0.05$, $\dS/\dx=1.4$, and $\Delta t \vortexstrengthq{1}/R_c^2=0.01$.}
    \label{fig:vortexnearcylinderpressure}
\end{figure}

\begin{figure}
    \centering
    \includegraphics[page=1]{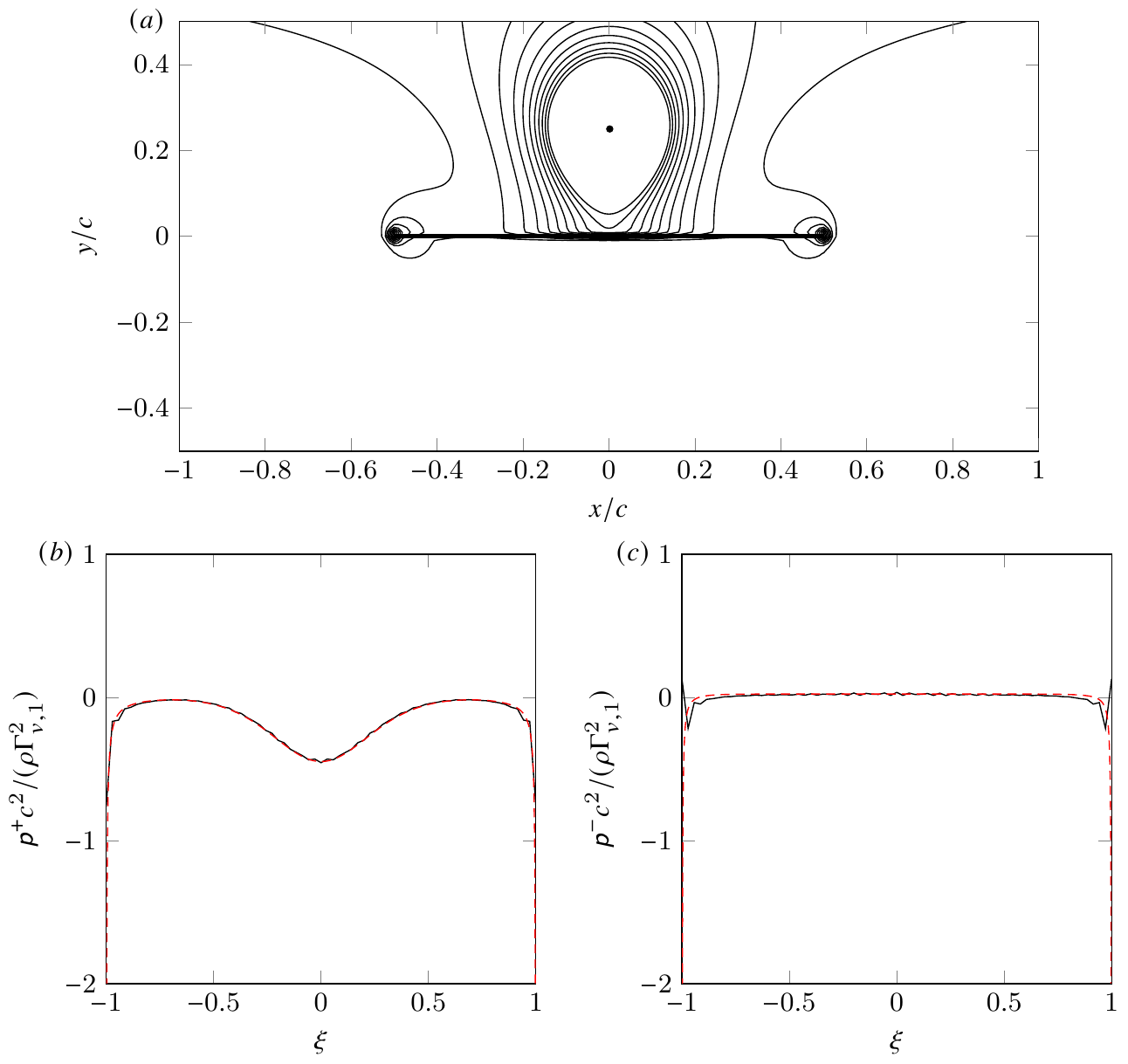}
    \caption{(\textit{a}) Contours (\full) of the discrete pressure for a point vortex (\fullcirc) with strength $\vortexstrengthq{1}$ at $(0,R_v)$ near a flat plate consisting of $\numpts$ points with chord length $c$ and a bound circulation $-\vortexstrengthq{1}$. The scaled discrete pressure (\full) at (\textit{b}) the top side and (\textit{c}) the bottom side of the plate. Overlaid is the exact continuous solution  (\textcolor{red}{\dashed}). The simulation is performed with  $R_v/c = 0.25$, $\dx/c=0.01$, $\dS/\dx=1.4$, and $\Delta t \vortexstrengthq{1}/c^2=0.005$.}
    \label{fig:vortexnearplatepressure}
\end{figure}

We now demonstrate the pressure calculations with two examples that have an analytical solution for the pressure distribution on the surface. Figure~\ref{fig:vortexnearcylinderpressure} and \ref{fig:vortexnearplatepressure} depict the pressure field and surface distribution for a vortex near a cylinder and a plate, respectively. The pressure inside the cylinder is close to the exact solution of a uniform value, except for some noise at the side near the vortex, which is visible on the interior surface distribution of the pressure. The exterior surface pressure distribution on the cylinder shows good agreement with the analytical solution. In the example of the flat plate, shown in figure~\ref{fig:vortexnearplatepressure}, the bottom and top surface pressure distributions again show good agreement with the analytical solution except for at the edges, which is expected. Note that in these examples, we used a smaller value for $\dS/\dx$ than before, which was necessary to prevent the low pressure from the vortex from leaking through the surface. As a result, high-frequency components of the surface pressure are incorrectly amplified through the regularization and interpolation operators~\cite{Goza2016}, especially at the edges.

\subsubsection{Impulse-based calculations of force and moment}

The integral of the pressure distribution over the surface (plus any edge-suction parameters in the case of sharp edges) will be equal to the force on the surface. However, in this section we provide an alternative means of calculating the force and moment on the body through the negative rate of change of impulse in the fluid. The continuous expressions for linear and angular impulse (about the origin) are, in two dimensions~\citep{Saffman1993,inviscidbook},
\begin{align}
    \boldsymbol{P} &= \int_{\domain} \x\times \boldsymbol{\omega}\,\mathrm{d}V + \int_{\surface} \x\times \left( \boldsymbol{n}\times\vel\right)\,\mathrm{d}S \label{eq:linimp}\\
    \boldsymbol{\Pi}_{\mathrm{O}} &= \frac{1}{2}\int_{\domain} \x\times \left(\x\times\boldsymbol{\omega}\right)\,\mathrm{d}V + \frac{1}{2} \int_{\surface} \x\times \left[ \x\times\left( \boldsymbol{n}\times\vel\right)\right]\,\mathrm{d}S \label{eq:angimp}
\end{align}
If there is only a single body, then the force and moment (about the origin) exerted by the fluid on that body are given by
\begin{equation}
\label{eq:FMimp}
    \boldsymbol{F} = -\rho \frac{\mathrm{d}\boldsymbol{P}}{\mathrm{d}t}, \qquad \boldsymbol{M}_{\mathrm{O}} = -\rho \frac{\mathrm{d}\boldsymbol{\angimp{}}_{\mathrm{O}}}{\mathrm{d}t}, 
\end{equation}
where $\rho$ is the fluid density. In the two-dimensional applications of this paper, the angular impulse and the moment have only a single component, e.g., $\boldsymbol{\angimp{}}_{\mathrm{O}} =\angimp{\mathrm{O}}\boldsymbol{e}_z$, where $\boldsymbol{e}_z$ is the unit vector out of the plane.

It should be observed that, by definition, the bound vortex sheet strength $\gamma$ is equal to the jump in tangential velocity between the fluid and the surface, $\boldsymbol{n}\times\vel= \gamma \boldsymbol{e}_z + \boldsymbol{n}\times\vel_{\body}$, where $\boldsymbol{n}$ is the unit surface normal vector directed into the fluid, $\vel$ is the fluid velocity, and $\vel_{b}$ is the velocity of the surface. Thus, the surface integrals in (\ref{eq:linimp}) and (\ref{eq:angimp}) can be re-written in terms of the vortex sheet strength and the body motion.

We can easily develop discrete forms of the integrals (\ref{eq:linimp}) and (\ref{eq:angimp}) with the solutions and notation described in this paper. For the volume integrals, let us denote diagonal matrices containing the coordinates of the grid nodes by $\diagmat{\xgrid}$ and $\diagmat{\ygrid}$. Thus, the expressions in (\ref{eq:linimp}) and (\ref{eq:angimp}) can be written in discrete form as
\begin{align}
    \linimp{x} &= \ipnodes{\ygrid}{\vortgrid} + \ipscalar{\ysurf}{\vsheet + \diagmat{\normvecc{x}} \velsurfc{y} - \diagmat{\normvecc{y}} \velsurfc{x} },\label{eq:linimpxd}\\
    \linimp{y} &= -\ipnodes{\xgrid}{\vortgrid} - \ipscalar{\xsurf}{\vsheet + \diagmat{\normvecc{x}} \velsurfc{y} - \diagmat{\normvecc{y}} \velsurfc{x} }, \label{eq:linimpyd}
\end{align}
and
\begin{equation}
\label{eq:angimpd}
\angimp{\mathrm{O}} = -\frac{1}{2} \ipnodes{\diagmat{\xgrid}\xgrid + \diagmat{\ygrid}\ygrid}{\vortgrid} -\frac{1}{2} \ipscalar{\diagmat{\xsurf}\xsurf + \diagmat{\ysurf}\ysurf}{\vsheet + \diagmat{\normvecc{x}} \velsurfc{y} - \diagmat{\normvecc{y}} \velsurfc{x}}.
\end{equation}

The overall force and moment exerted on the body are obtained from calculating these impulses and computing their rates of change in (\ref{eq:FMimp}). Part of this force and moment is attributable to the dynamics of vorticity in the fluid. The remaining part is due to surface motion relative to the fluid, and we will discuss this in the next section.

\begin{figure}
    \centering
    \includegraphics[page=1]{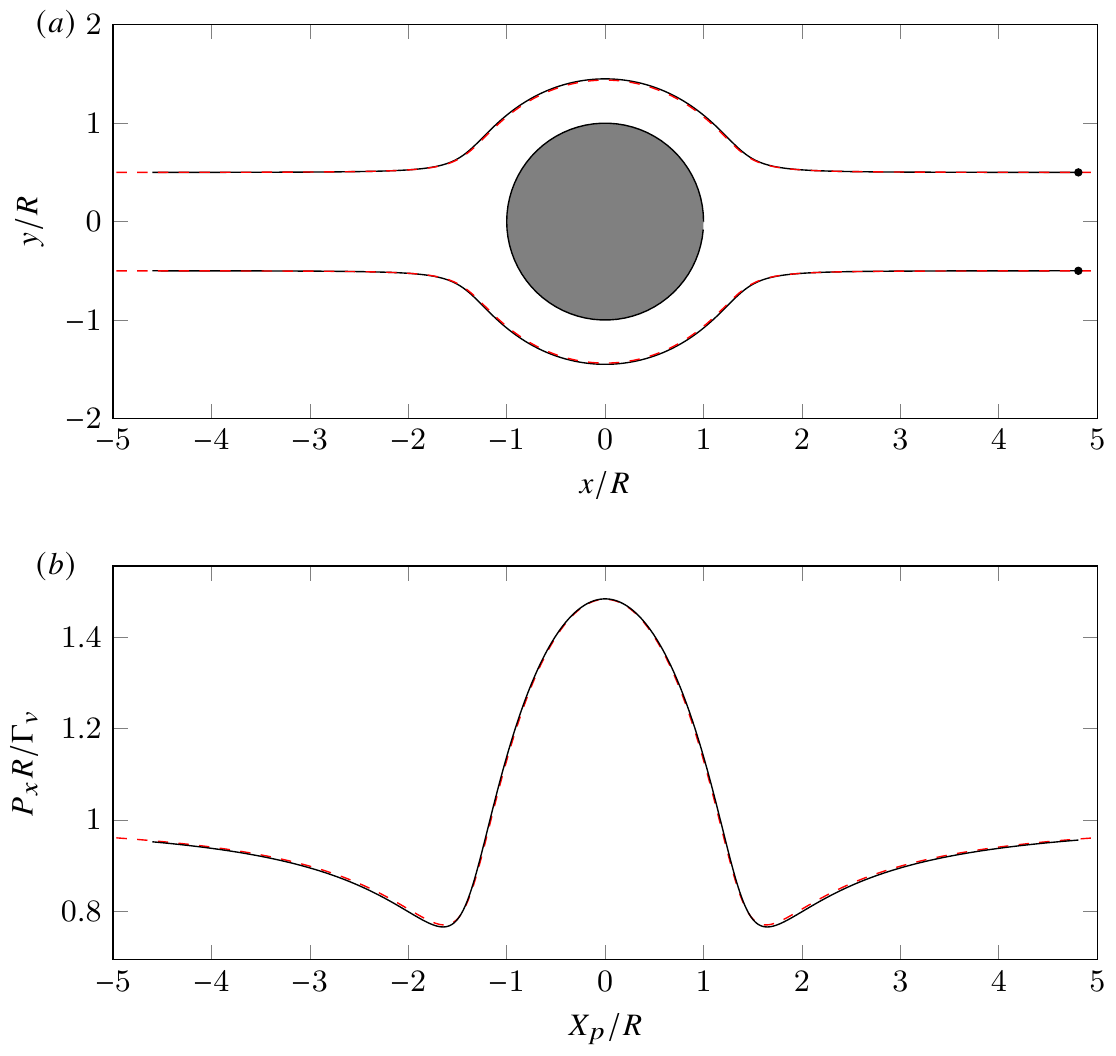}
    \caption{(\textit{a}) Numerically simulated trajectories (\full) of two point vortices (\fullcirc) of opposite strengths $\vortexstrengthq{1}$ and $\vortexstrengthq{1}=-\vortexstrengthq[2]$ being convected past a circular cylinder with radius $R$, and (\textit{b}) the $x$ component of the associated, numerically simulated impulse (\full) in the fluid. Overlaid are the exact continuous trajectories  and impulse (\textcolor{red}{\dashed}). The simulation is performed with $\dx/R=0.04$, $\dS/\dx=2$, and $\Delta t \vortexstrengthq{1}/R^2=0.1$.}
    \label{fig:impulse}
\end{figure}

\begin{figure}
    \centering
    \includegraphics[page=1]{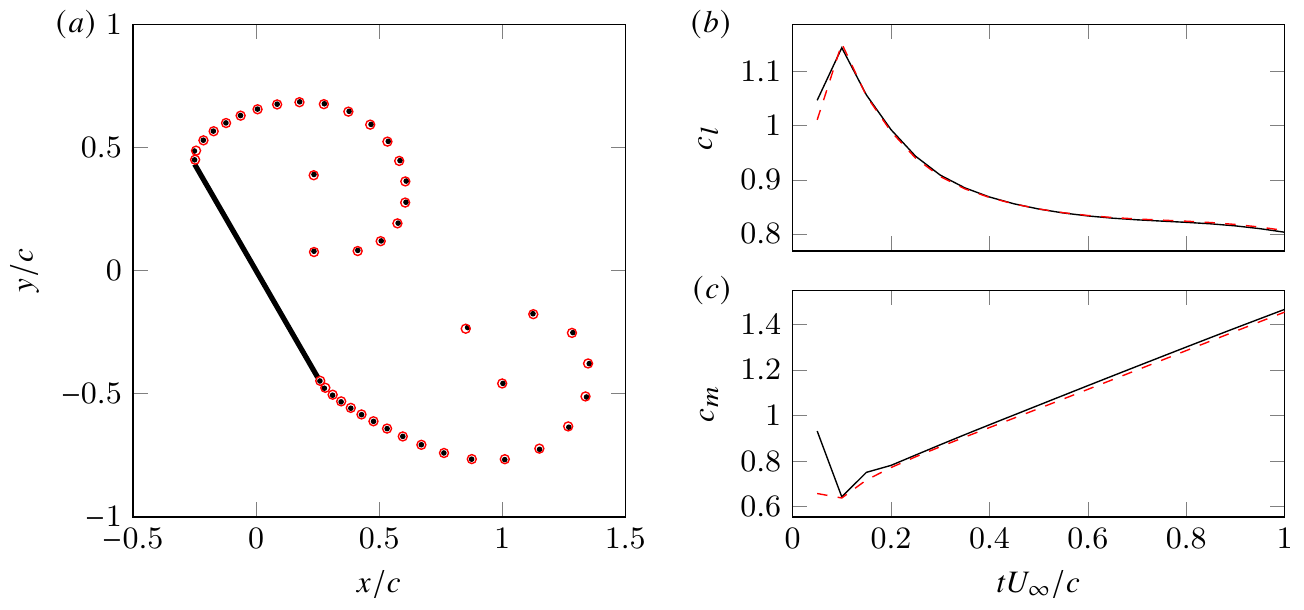}
    \caption{Comparison of the simulated vortex shedding behind a flat plate of length $c$ at \SI{60}{\degree} in a uniform flow using the method in this paper (\fullcirc\,and \full) and using the Biot-Savart method of \citet{darakjde2018} (\textcolor{red}{\opencirc}\,and \textcolor{red}{\dashed}). (\textit{a}) The positions of the shedded point vortices, (\textit{b}) the lift coefficient, and (\textit{c}) the moment coefficient, one convective time after impulsively starting the uniform flow. At both edges, the Kutta condition is enforced. The simulation is performed with $\dx/c=0.01$, $\dS/\dx=2$, and $\Delta t U_{\infty}/c=0.05$.}
    \label{fig:vortexsheddingcomparison}
\end{figure}

To illustrate the accuracy of the impulse-based calculation of force, we apply the method to two examples. In the first example, we simulate the trajectories of two point vortices of opposite strength, in which case each vortex is convected past a cylinder due to the presence of the other vortex. The time stepping in this example is carried out using a fourth-order Runge-Kutta scheme. Figure~\ref{fig:impulse} shows the trajectories and the $x$ component of the impulse together with their exact solutions, which show good agreement with the exact solution. In the second example, we compare our simulation of the first instants of the unsteady, fully separated flow around a flat plate after impulsively starting a uniform flow with the Biot-Savart method from \citet{darakjde2018}, using the same positioning rules to insert point vortices and the same time step. The vortex positions and the corresponding impulse and lift are compared in figure~\ref{fig:vortexsheddingcomparison} and show good agreement as well.

\subsubsection{Added mass}

The added mass tensor provides a measure of the inertial influence of the fluid on the body in response to changes in the body's translational or rotational motion. The coefficients of the added mass tensor of a body are obtained by computing the impulse components associated with a unit-valued component of motion. The motion's influence is both direct, via the surface velocity, and indirect, in the bound vortex sheet that develops on the surface.

For example, suppose that we consider translation at unit velocity in the $x$ direction, for which the motion is described by $\velsurfc{x} = \onesspoint$, $\velsurfc{y} = 0$, and $\sfsurfp = \ysurf$, and the associated bound vortex sheet---obtained without the Kutta condition by solving the basic problem (\ref{eq:basicblock})---is $\vsheet = \schurA^{-1}\projectC\sfsurfp = \schurA^{-1}\projectC\ysurf$. The added mass coefficients corresponding to this motion are derived by substituting these into the impulse formulas (\ref{eq:linimpxd})--(\ref{eq:angimpd}):
\begin{align}
\linimp{x}^{(x)} &= \ipscalar{\ysurf}{\schurA^{-1}\projectC\ysurf - \diagmat{\normvecc{y}}\onesspoint}  \\ 
\linimp{y}^{(x)} &= -\ipscalar{\xsurf}{\schurA^{-1}\projectC\ysurf - \diagmat{\normvecc{y}}\onesspoint}, \\
\angimp{\mathrm{O}}^{(x)} &= -\frac{1}{2} \ipscalar{\diagmat{\xsurf}\xsurf + \diagmat{\ysurf}\ysurf}{\schurA^{-1}\projectC\ysurf - \diagmat{\normvecc{y}} \onesspoint}.
\end{align}
Thus, the components of the added mass coefficients tensor associated with translation in the $x$ direction are
\begin{align}
m^{F}_{xx} &= \rho \linimp{x}^{(x)}\\
m^{F}_{xy} &= \rho \linimp{y}^{(x)},\\
m^{M}_{x} &= \rho \left(\angimp{\mathrm{O}}^{(x)}-\xcentvec \times \boldsymbol{P}^{(x)} \right),
\end{align}
where $\xcentvec$ is the centroid of the body, which can be calculated using (\ref{eq:centroid}), and the superscript $F$ and $M$ are used to denote the coefficient for the force and moment, respectively.

A similar approach can be used to obtain the added mass coefficients due to unit translation in the $y$ direction, for which $\velsurfc{x} = 0$, $\velsurfc{y} = \onesspoint$, and $\sfsurfp = -\xsurf$. The coefficients due to unit rotation follow from taking $\velsurfc{x} = -\ysurf$, $\velsurfc{y} = \xsurf$, and $\sfsurfp = -\frac{1}{2}(\diagmat{\xsurf}\xsurf+\diagmat{\ysurf}\ysurf)$.

\subsection{Multiple bodies}
\label{sec:multibody}

The previous sections provided the formulations for potential flow with the presence of a body. The extension of these expressions to multiple bodies is straightforward and consists of allocating partitions of $\vsheet$ to the different bodies. The surface streamfunction has to be partitioned accordingly, with the body motion streamfunction $\sfsurf$ containing the values for the discrete surface points from all the bodies and $\sfzero = \sum_j \onesspoint_j \szeroj{j}$ allocating a uniform value $\szeroj{j}$ to the $j$th body, where $\onesspoint_j \in \spoints$ is a vector whose $i$th component is one if it belongs to the $j$th body and zero otherwise. The system (\ref{eq:basicblock}) can then be solved for the streamfunction field without modification.

\begin{figure}
    \centering
    \includegraphics[page=1]{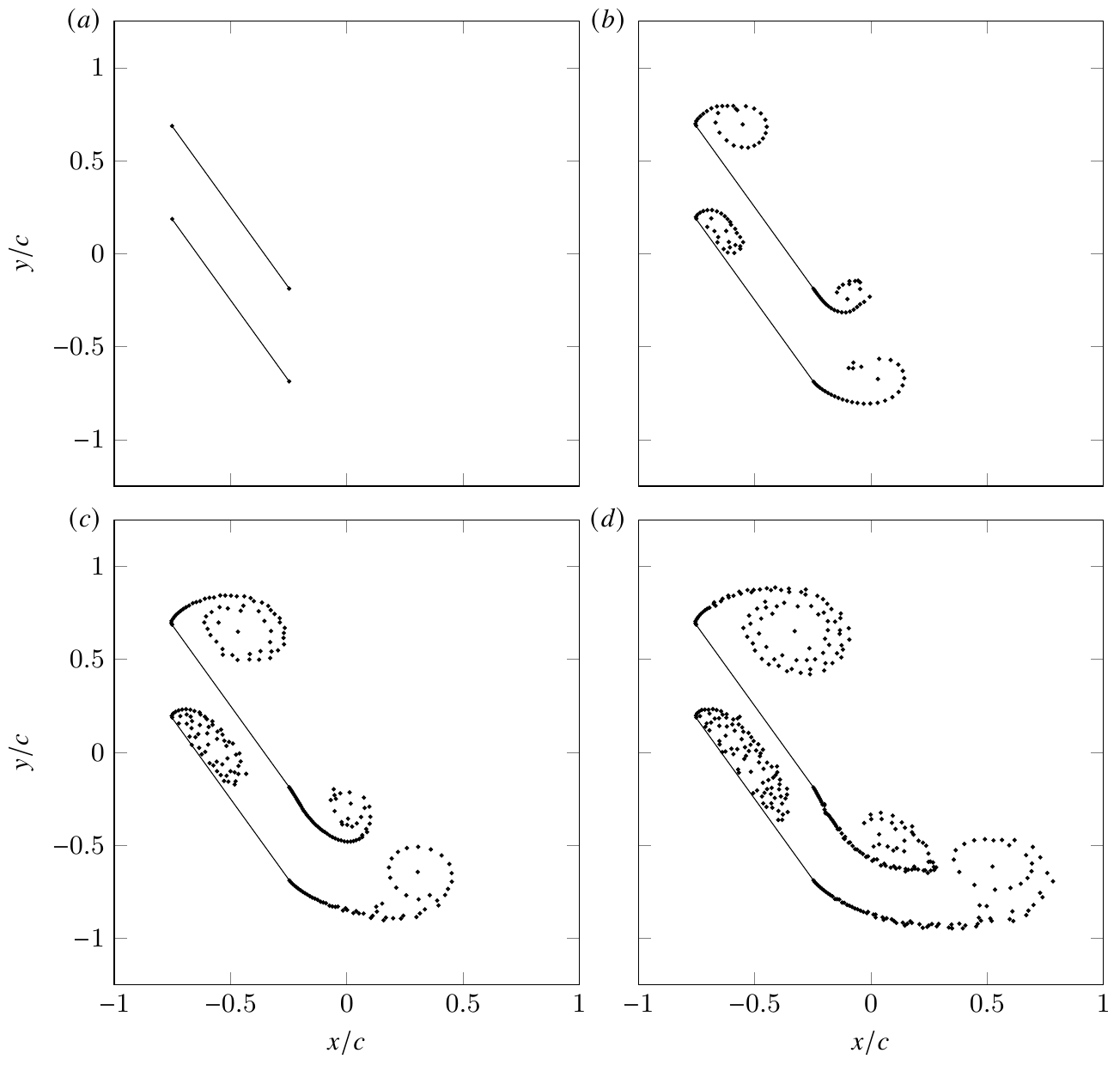}
    \caption{Evolution of the flow around two flat plates of length $c$ at \SI{60}{\degree}, vertically separated by a distance of $c/2$, at $t U_{\infty}/c$ equal to (\textit{a}) $0$, (\textit{b}) $0.3$, (\textit{c}) $0.6$, and (\textit{d}) $0.9$ after impulsively starting a uniform flow. The simulation is performed with $\dx/c=0.01$, $\dS/\dx=2$, and $\Delta t U_{\infty}/c=0.01$.}
    \label{fig:twoplates}
\end{figure}

As in the single-body case, if we want to enforce an edge condition on the $j$th body, we treat its uniform streamfunction value $\szeroj{j}$ as a Lagrange multiplier and add a constraint on $\vsheet$ to the saddle point system. In the unsteady case, we add a circulation constraint for each body and consider each new point vortex to be released from a specified body. For example, let us consider two bodies, with each body having one sharp edge. We assume $\vortcirc{1}$ and $\vortcirc{2}$ are the strengths from the vortices that were released from the first and second body, respectively. We can compute these strengths by enforcing the Kutta condition for both bodies using the saddle point system
\begin{equation}
\label{eq:kuttasystem4}
\begin{bmatrix}
\lapgrid & \regdssmooth & 0 & 0 & \unitgridvort{1} & \unitgridvort{2} \\
\interp & 0 & \onesspoint_1 & \onesspoint_2 & 0 & 0\\
0 & \unitspoint{k_{1}}^{T}  & 0 & 0 & 0 & 0 \\
0 & \unitspoint{k_{2}}^{T}  & 0 & 0 & 0 & 0 \\
0 & \vsheetzeroopj{1} & 0 & 0 & 1 & 0 \\
0 & \vsheetzeroopj{2} & 0 & 0 & 0 & 1
\end{bmatrix}
\begin{pmatrix}
\sfgrid \\
\svsheet \\
\szeroj{1} \\
\szeroj{2} \\
\vortcirc{1} \\
\vortcirc{2} 
\end{pmatrix} = 
\begin{pmatrix}
-\vortgrid \\
\projectC\sfsurfp  \\
0 \\
0 \\
-\circwj{1} \\
-\circwj{2}
\end{pmatrix},
\end{equation}
where we defined $\circwj{j}$ as the circulation of the existing vorticity in the flow that has previously been released from the $j$th body and $\vsheetzeroj{j} = \diagmat{\vsheetzero}\onesspoint_j$. Again, these solutions are easily extensible to an arbitrary number of edges per body. For example, Figure~\ref{fig:twoplates} demonstrates the method for two flat plates in a uniform flow where the LE and TE are regularized for both plates by releasing four point vortices during each time step.

\begin{figure}
    \centering
    \includegraphics[page=1]{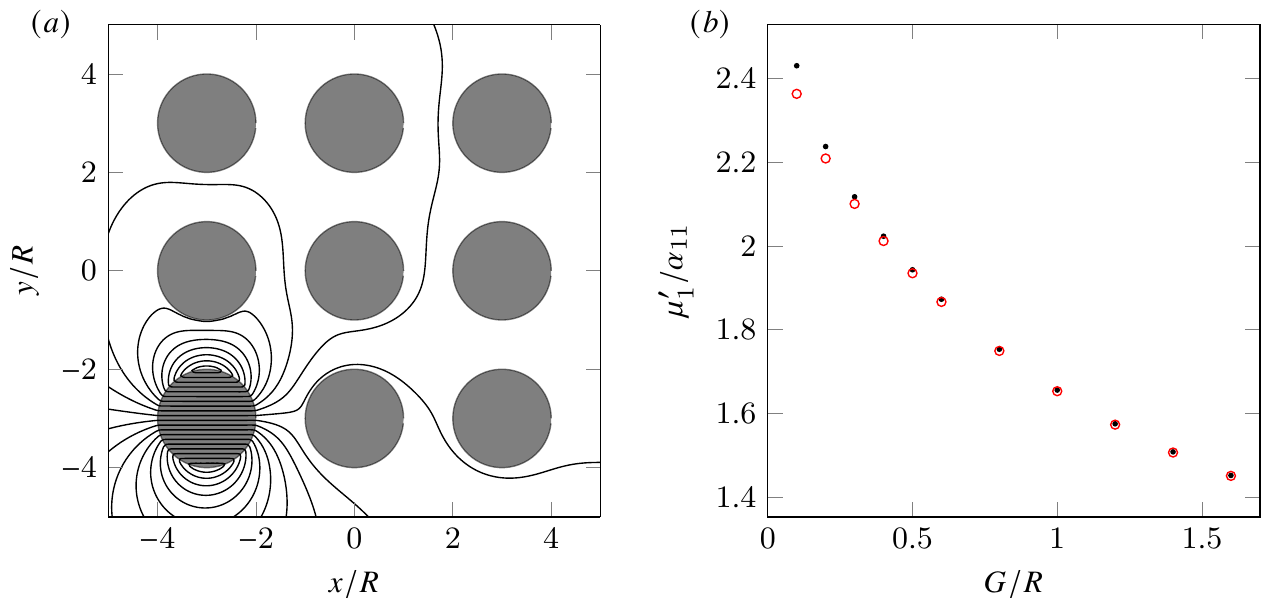}
    \caption{(\textit{a}) Contours of the discrete streamfunction for an array of nine circular cylinders with radius $R$, spaced with a gap distance $G$ between each cylinder, of which the bottom left cylinder translates horizontally. (\textit{b}) Numerically simulated variation (\fullcirc) of the ratio between the largest eigenvalue of the added mass coefficient tensor and the largest self-added mass coefficient with the gap-to-radius ratio $G/R$. Overlaid are the values (\textcolor{red}{\opencirc}) obtained by \citet{Chen1975} from solving a system of truncated analytical expressions. The simulations are performed with $\dx/R=0.05$ and $\dS/\dx=2$.}
    \label{fig:cylinderarray}
\end{figure}

The formulas for pressure, force and added mass also generalize to systems with multiple bodies. For example, figure \ref{fig:cylinderarray} demonstrates an example of a potential flow model with an array of nine circular cylinders and compares the ratio of the largest eigenvalue of added mass coefficient tensor and the largest self-added mass coefficient of the system with the results of \citet{Chen1975}.

\section{Conclusion}

A treatment of potential flow on Cartesian grids was presented. The main body of this work is based on the computation of the discrete streamfunction through the streamfunction-vorticity Poisson equation with singular vortex elements as vorticity sources. The potential flow in the presence of sinks and sources requires the computation of the discrete scalar potential, which is completely analogous except for the scalar potential Nuemann boundary condition instead of the streamfunction Dirichlet boundary condition to enforce no-penetration on surfaces in the flow. The Helmholtz decomposition then shows how the velocity fields associated with the scalar potential and streamfunction can be superposed to obtain the combined flow due to sources, sinks, vortices.

In our potential flow treatment, we used two algebraic techniques that allowed us to mimic the analytical treatment of potential flows around sharp-edged bodies with bound vortex sheets. The first technique is to account for surfaces in the flow by using the immersed boundary projection method. We introduce a Lagrange multiplier for the no-penetration constraint in the streamfunction-vorticity Poisson equation and identify it as a discrete version of the continuous strength distribution of a bound vortex sheet on the body. The discrete equations that solve the associated saddle-point system are then completely analogous to the continuous boundary integral equations. It should be noted that this underlying continuous problem is a Fredholm integral equation of the first kind and is ill-posed, leading to poorly-conditioned discrete operators in the solution method. We have not attempted to address this issue here, but the work of~\citet{Goza2016} did and its methods are straightforward to apply. The second algebraic technique is to decompose the discrete bound vortex sheet strength for sharp-edged bodies into a singular and non-singular part. One can then add constraints on the elements of the non-singular part that are located at the edges to make the recomposed bound vortex sheet well-behaved. This way, we enforced the Kutta condition in a way that is similar to analytical treatments of the Kutta condition and that allows for generalized edge conditions as well. Furthermore, we leveraged the concept of the discrete bound vortex sheet strength to create expressions for the pressure in the flow and on surfaces, the impulse in the flow around surfaces, and the added mass matrix for a system of arbitrarily shaped bodies. Finally, it is important to note that the treatment of potential flow that we presented is not restricted to the specific finite-difference discretization tools that we used in our implementation to provide the examples in this work. Also, one is not restricted to use the immersed-boundary projection method to obtain an expression for the discrete vortex sheet strength and can use, for example, the immersed-interface method instead as in~\cite{Gillis2019}.

We found that the method can accurately replicate the results of a Biot-Savart method for the unsteady flow around a flat plate. This motivates the goal of implementing a three-dimensional version of this method using vortex particles (with a vector representing their strengths) or filaments~\citep{Cottet2000}, since the concepts of the immersed boundary projection method and the enforcement of edge conditions through the multiplicative decomposition of the vortex sheet strength generalize to three dimensions. Such a three-dimensional, grid-based solver would rely on the vector potential and a vector treatment of the vorticity field, while properly accounting for vortex stretching. It could potentially make significant cost improvements over unsteady three-dimensional panel methods, which generally scale poorly as the number of panels and vortex elements in the flow increase.

\section*{Acknowledgements}
The support for this work by the US Air Force Office of Scientific Research (FA9550-18-1-0440) with programme manager Gregg Abate is gratefully acknowledged.


\appendix
\section{Solution of general saddle-point systems}
\label{app:saddlesystems}

A general block system (with positive semi-definite matrix $\basicmat$) can be decomposed as follows:
\begin{equation}
\begin{bmatrix}
\basicmat & \basicBoneT \\ \basicBtwo & -\basicC
\end{bmatrix} = \begin{bmatrix}
\basicmat & 0\\ \basicBtwo & \basicS
\end{bmatrix}
\begin{bmatrix}
\grid{I} & \basicmat^{-1}\basicBoneT\\ 0 & \grid{I}
\end{bmatrix},
\end{equation}
where
\begin{equation}
\label{eq:schur}
\basicS \equiv -\basicC - \basicBtwo\basicmat^{-1}\basicBoneT
\end{equation}
is the {\em Schur complement} of the matrix system and $\grid{I}$ is the identity. By this decomposition, we can develop an algorithm for the solution of the block system
\begin{equation}
\label{eq:saddlept}
\begin{bmatrix}
\basicmat & \basicBoneT \\ \basicBtwo & -\basicC
\end{bmatrix} \begin{pmatrix} \basicx \\ \basicy \end{pmatrix} = \begin{pmatrix} \basicrone \\ \basicrtwo \end{pmatrix}.
\end{equation}
We will refer to $\basicx$ as the solution vector and $\basicy$ as the constraint force. We define the intermediate solution vector $(\basicx^{*},\, \basicy^{*})^{T}$ as the solution of the lower-triangular system
\begin{equation}
\begin{bmatrix}
\basicmat & 0\\ \basicBtwo & \basicS
\end{bmatrix} \begin{pmatrix} \basicx^{*} \\ \basicy^{*} \end{pmatrix} = \begin{pmatrix} \basicrone \\ \basicrtwo \end{pmatrix} 
\end{equation}
and then the solution we seek can be found by back substitution of
\begin{equation}
\begin{bmatrix}
\grid{I} & \basicmat^{-1}\basicBoneT\\ 0 & \grid{I}
\end{bmatrix} \begin{pmatrix} \basicx \\ \basicy \end{pmatrix} = \begin{pmatrix} \basicx^{*} \\ \basicy^{*} \end{pmatrix} 
\end{equation}
The algorithm we derive from this is
\begin{align}
\basicmat \basicx^{*} &= \basicrone, \nonumber\\
\basicS \basicy^{*} &=  \basicrtwo - \basicBtwo \basicx^{*}, \label{eq:algor}\\
\basicy &= \basicy^{*}, \nonumber\\
\basicx &= \basicx^{*} - \basicmat^{-1}\basicBoneT \basicy \nonumber.
\end{align}
It is also useful to have an inverse representation of the block matrix system:
\begin{equation}
\label{eq:saddleinverse}
\begin{pmatrix} \basicx \\ \basicy \end{pmatrix} = \begin{bmatrix} \basicmat^{-1} + \basicmat^{-1} \basicBoneT\basicS^{-1} \basicBtwo \basicmat^{-1} & -\basicmat^{-1}\basicBoneT \basicS^{-1} \\ -\basicS^{-1} \basicBtwo \basicmat^{-1} & \basicS^{-1}\end{bmatrix} \begin{pmatrix} \basicrone \\ \basicrtwo \end{pmatrix}. 
\end{equation}

\section{Some geometric relations for discrete surfaces}

Consider a closed surface $\surfb$ with unit normal $\nrm$. We will recall some basic geometric relations here, and then provide some discrete versions of these relations based on the set of points with coordinates $\xsurf$, $\ysurf$, normal components $\normvecc{x}$ and $\normvecc{y}$ (which, the reader will recall, contain the surface length or area of each segment or panel associated with the points).

The volume $\volb$ of the region enclosed by $\surfb$ can be computed from the integral
\begin{equation}
    \volb = \frac{1}{n_d} \int_{\surfb} \x\cdot \nrm\,\mathrm{d}S,
\end{equation}
where $n_d$ is the number of spatial dimensions (2 or 3). Using the notation above, the approximate form of this expression is
\begin{equation}
    \volb \approx \frac{1}{n_d} \sum_j \rsurf{j}^T \normvecc{j},
\end{equation}
where the sum is taken over the $n_d$ components.

An alternative formula for the volume is
\begin{equation}
    \volb \boldsymbol{e}_j = -\frac{1}{n_d-1}\int_{\surfb}\x\times(\boldsymbol{n}\times \boldsymbol{e}_j)\,\mathrm{d}S.
\end{equation}
The components of this integral can be written discretely as
\begin{equation}
    \volb \boldsymbol{e}_j \approx \frac{\boldsymbol{e}_j}{n_d-1} \sum_{k\neq j} \rsurf{k}^T \normvecc{k}
\end{equation}

And finally, a third alternative is 
\begin{equation}
    \volb \boldsymbol{I} = \int_{\surfb} \x\nrm\,\mathrm{d}S,
\end{equation}
where $\boldsymbol{I}$ is the identity. The discrete form of this is a diagonal matrix with $\rsurf{x}^T\normvecc{x}$, $\rsurf{y}^T\normvecc{y}$, and $\rsurf{z}^T\normvecc{z}$ along the diagonal. 

Thus, we can conclude that the volume of the body is approximately
\begin{equation}
    \volb \approx \ipscalar{\rsurf{x}}{\normvecc{x}} \approx \ipscalar{\rsurf{y}}{\normvecc{y}} \approx \ipscalar{\rsurf{z}}{\normvecc{z}},
\end{equation}
or any average of some combination of these.

The centroid of the body can be derived from the equation
\begin{equation}
    \xcentvec \volb
= \frac{1}{2} \int_{\surfb} \x\cdot\x\boldsymbol{n}\,\mathrm{d}S,
\end{equation}

or, in discrete form,
\begin{equation} \label{eq:centroid}
    \xcent \approx \frac{1}{2\volb}\ipscalar{\diagmat{\xsurf} \xsurf + \diagmat{\ysurf} \ysurf}{\normvecc{x}},\qquad \ycent \approx \frac{1}{2\volb}\ipscalar{\diagmat{\xsurf} \xsurf + \diagmat{\ysurf} \ysurf}{\normvecc{y}}.
\end{equation}

\bibliographystyle{jfm}
\bibliography{main}

\end{document}